**Title: A synthetic data–enabled artificial intelligence system for pan-tumour CT screening and diagnosis: a retrospective simulation study**

Wenhui Lei[1,2*], Hanyu Chen[3*], Zitian Zhang[4*], Luyang Luo[5*], Qiong Xiao[3], Yannian Gu[1], Peng Gao[3], Yankai Jiang[2], Ci Wang[4], Guangtao Wu[3], Tongjia Xu[3], Yingjie Zhang[3], Pranav Rajpurkar[5], Xiaofan Zhang[1,2#], Shaoting Zhang[1#], Zhenning Wang[3#]

[1]Shanghai Jiao Tong University, Shanghai, P.R. China
[2]Shanghai AI Lab, Shanghai, P.R. China
[3]Department of Surgical Oncology and General Surgery, Key Laboratory of Precision Diagnosis and Treatment of Gastrointestinal Tumours, Ministry of Education, The First Hospital of China Medical University, Liaoning, P.R. China
[4]Department of Radiology, The First Hospital of China Medical University, Liaoning, P.R. China
[5]Harvard Medical School, Boston, MA, USA

*Contributed equally as first authors
#Correspondence to Xiaofan Zhang, Shaoting Zhang or Zhenning Wang

**Summary**

**Background**

Artificial intelligence (AI) has advanced oncology imaging, but progress towards generalisable systems for pan-tumour CT interpretation is constrained by limited access to large-scale, lesion-level annotated datasets, driven by privacy restrictions, annotation cost, and the rarity of many tumour types. We aimed to develop and evaluate a synthetic data-enabled AI system to support radiologists in pan-tumour CT screening and diagnosis.

**Methods**

We developed PASTA, a lesion-aware model trained using PASTA-Gen, a synthetic data framework that generates 3D CT scans with pixel-level lesion masks and structured lesion descriptions across ten organ systems. Using PASTA-Gen, we constructed PASTA-Gen-30K, comprising 30,000 synthetic 3D CT image-mask-text pairs spanning malignant and benign lesions. We assessed model performance across multiple oncology imaging tasks and evaluated clinical utility by integrating the model into a workflow-aligned decision-support system (PASTA-AID). We conducted a retrospective simulated clinical study including two scenarios: time-limited non-contrast CT tumour screening and contrast-enhanced CT diagnosis workflows for lesion segmentation and structured report preparation.

**Findings**

Across a broad range of downstream oncology tasks, PASTA showed consistently strong performance and generalisation, supporting non-contrast CT tumour screening, lesion segmentation, structured report generation, staging, survival-related prediction, and cross-modality transfer. In simulated time-limited, high-workload non-contrast CT screening, PASTA-AID improved radiologists' throughput by 11.1–25.1%, increased sensitivity by 17.0–31.4%, and increased precision by 10.5–24.9%. Additionally, in diagnosis workflows on contrast-enhanced CT, PASTA-AID reduced lesion segmentation time by up to 78.2% and structured report preparation time by up to 36.5%, while improving agreement between less-experienced and senior radiologists.

**Interpretation**

This work establishes an end-to-end synthetic data–driven pipeline encompassing data generation, model development, and clinical validation, demonstrating the value of synthetic data for scalable pan-tumour CT analysis and workflow-oriented clinical decision support.

**Funding**

National Natural Science Foundation of China, Shanghai Municipal Commission of Economy and Informatization, National Science and Technology Major Project

## Introduction

Malignant tumours remain one of the leading causes of death worldwide, placing substantial medical and economic burdens on healthcare systems[1]. Radiological examination, exemplified by computed tomography (CT), plays a key role in oncology by supporting tumour detection, staging, treatment planning, response assessment, and follow-up. In recent years, artificial intelligence (AI) has been increasingly integrated into radiological image interpretation, with the aim of improving diagnostic accuracy, efficiency and reproducibility. Task-specific AI models have demonstrated strong performance in organ- and tumour-focused applications, including detection, segmentation and prognostic prediction[2-7]. Nevertheless, these models remain constrained by limited disease coverage, heavy dependence on annotated datasets and insufficient generalisation across tumour types and clinical settings.

Foundation models (FMs), pre-trained on large, heterogeneous datasets, have emerged as a powerful paradigm for medical imaging, offering improved label efficiency and robustness across diverse downstream tasks[8-16]. Nevertheless, developing a pan-tumour radiology FM remains challenging, with several obstacles contributing to this gap. On cross-sectional imaging such as CT, tumours often occupy only a small fraction of the volume, and most existing radiology FMs are trained with self-supervised objectives on unlabelled scans[17-20]. Such objectives tend to emphasize global anatomical structure rather than subtle lesion-specific signals, which limits utility for oncology imaging. Recent efforts toward supervised foundation models, exemplified by SuPreM[21], demonstrate the feasibility of incorporating labeled anatomical or lesion information during pre-training and demonstrate improved attention to target-region features relative to purely self-supervised counterparts. However, the scarcity of high-quality supervised resources, particularly lesion-level labels and organ-specific annotations across tumour types, restricts the scale of pre-training and narrows tumour coverage, which in turn constrains generalisability. Finally, the development of large, multi-tumour datasets is impeded by the intrinsic rarity of many tumours, the cost of expert annotation, and privacy constraints.

To address these challenges, we developed a synthetic data-enabled artificial intelligence system designed to support pan-tumour CT screening and diagnosis. Using large-scale lesion-level synthetic data, the system learns tumour-relevant imaging patterns that generalise across organs and clinical tasks. We assessed its performance across a range of oncologic imaging applications and examined its clinical utility by embedding it into a workflow-aligned decision-support system evaluated through a retrospective simulated clinical study covering time-limited screening and diagnostic workflows. This work presents a synthetic data-enabled approach for building scalable artificial intelligence systems for oncology imaging that are aligned with real-world clinical workflows.

## Methods

### Study design and participants

This was a multi-stage, retrospective, simulation-based methodological study to develop and evaluate a clinical decision-support framework enabled by large-scale pre-trained AI models for oncologic imaging. The study involved no prospective recruitment and no real-time clinical deployment.

For synthetic lesion generation, we retrospectively collected two institutional datasets from The First Hospital of China Medical University (CMU), China (figure 1). First, 10 767 de-identified CT scans acquired between Jan 1, 2022, and Dec 31, 2023, were randomly sampled as template scans. These scans included both non-contrast and contrast-enhanced studies across multiple anatomical regions and were paired with corresponding radiology reports (appendix p24-25). Second, reference lesion data for simulation were collected from CT scans acquired between Jan 1, 2020, and Dec 31, 2023, comprising 750 cases covering 15 lesion types, with no individual overlap with the template cohort (CMU-LSR, appendix p26).

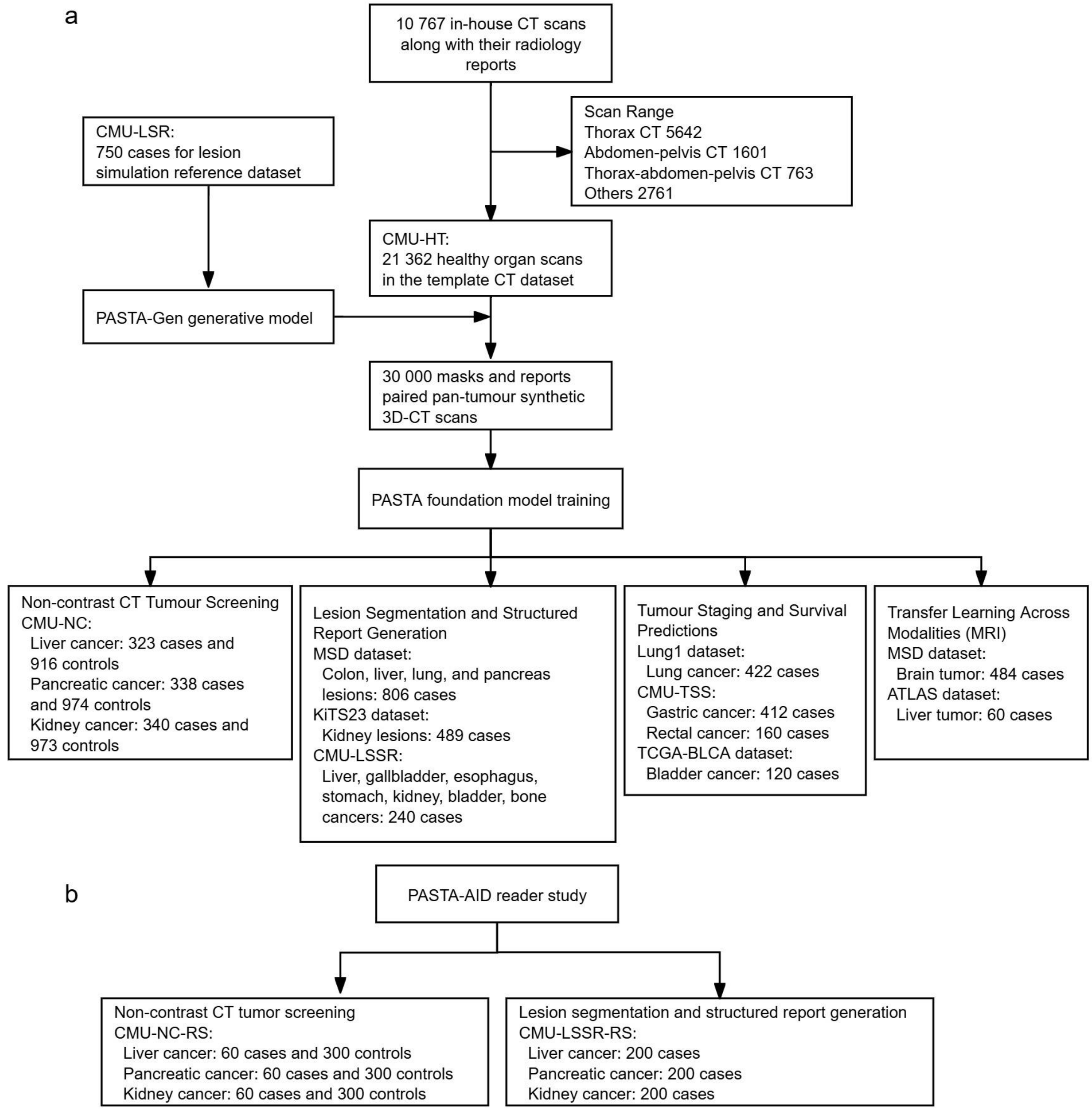

**Figure 1. Data sources and cohort composition**

**a**, Overview of data sources used for synthetic dataset construction, model training, and downstream evaluations.
**b**, Cohorts used in the retrospective simulated reader study.
CMU = The First Hospital of China Medical University
CMU-HT = China Medical University Healthy Template dataset
CMU-LSR = China Medical University Lesion Simulation Reference dataset
CMU-NC = China Medical University Non-contrast CT dataset
CMU-LSSR = China Medical University Lesion Segmentation and Structured Report Generation Datasets
CMU-TSS = China Medical University Tumour Staging and Survival cohort

MRI = Magnetic resonance imaging
CT = Computed tomography

For model validation tasks, including non-contrast CT cancer screening, lesion segmentation, structured report generation, and cross-modal generalisation, a combination of institutional and public datasets was used. For non-contrast CT cancer screening, CT scans of liver, pancreatic, and kidney cancers with matched controls were retrospectively collected from The First Hospital of CMU between Jan 1, 2020, and Dec 31, 2023 (n=3864, CMU-NC), including 323 liver cancer cases with 916 controls, 338 pancreatic cancer cases with 974 controls, and 340 kidney cancer cases with 973 controls. For lesion segmentation, public datasets included the Medical Segmentation Decathlon[22] (MSD; n=1290), comprising multi-institutional cohorts from Europe and North America (including the University of Pennsylvania, IRCAD Hôpitaux Universitaires, The Cancer Imaging Archive, and Memorial Sloan Kettering Cancer Center) and covering brain tumours, liver tumours, lung tumours, pancreatic tumours, colon tumours, and pancreas cysts. Additional public datasets included the KiTS23[23] dataset (n=489), comprising kidney tumour and cyst CT scans sourced from M Health Fairview, North America, and the ATLAS dataset from the University Hospital of Dijon, France (n=60). We additionally included institutional CT scans collected at The First Hospital of CMU between Jan 1, 2020, and Dec 31, 2023, for segmentation of gallbladder cancer, esophageal cancer, gastric cancer, bladder cancer, bone metastases, liver cysts, gallstones, and kidney stones (n=240, CMU-LSSR, appendix p27). For tumour staging and survival prediction, evaluations were performed using the Lung1 dataset[24] (n=422; MAASTRO Clinic, the Netherlands), TCGA-BLCA[25] (n=120; multi-institutional cohort from The Cancer Genome Atlas), and additional institutional cohorts of gastric and rectal cancers collected at The First Hospital of CMU between Jan 1, 2020, and Dec 31, 2023 (572 cases, CMU-TSS).

For the retrospective simulated reader study, we collected independent institutional CT datasets from The First Hospital of CMU between Jan 1, 2015, and Dec 31, 2021 for three clinical tasks: non-contrast CT tumour screening, lesion segmentation, and structured report generation. For non-contrast CT tumour screening, we included 60 tumour cases and 300 control scans per cancer type (liver, pancreas, and kidney), resulting in 1080 non-contrast CT scans in total (CMU-NC-RS). For lesion segmentation and structured report generation, we collected contrast-enhanced CT scans for the same three cancer types (200 cases per cancer type; n=600 total, CMU-LSSR-RS). Reader-study datasets were used exclusively for evaluation and were not included in model fine-tuning.

The simulated reader study involved four radiologists (two junior and two senior) from The First Hospital of CMU who interacted with the system solely for evaluation using retrospective cases. Reader interactions did not influence patient care or clinical management.

**Synthetic lesion generation, quality evaluation, and dataset development**

Synthetic lesions were generated using PASTA-Gen, a rule-guided generative framework that models lesion appearance based on statistical analysis of real reference lesions and iterative input from experienced radiologists. Lesion characteristics were parameterised using eight structured attributes capturing clinically relevant variability, including CT enhancement status, size, morphology, density, heterogeneity, boundary characteristics, surface features, and invasion behaviour (appendix p23). Lesion size was sampled from log-normal distributions, while morphology and appearance were adapted to lesion- and organ-specific patterns observed in both solid and hollow organs. Density and heterogeneity were modelled relative to surrounding normal tissues, whereas invasion and surface were parameterised as no close versus close relationship with adjacent structures and well-defined versus ill-defined margins, respectively.

Template CT scans were standardised, organ-segmented, and filtered to ensure lesion simulation on anatomically healthy organs. A three-dimensional denoising diffusion model was applied to refine the initially synthesised lesions and suppress visual artefacts (appendix p18).

To assess synthetic data quality, we conducted a blinded reader study involving four radiologists. Image realism was evaluated using a blinded Turing-style assessment in which real and synthetic scans were randomly mixed at a 1:1 ratio (n=1180) and rated on a five-point Likert scale (appendix p19). Text–image consistency was assessed on 750 synthetic image–mask–text pairs across five predefined attributes (shape, density, heterogeneity, surface, and invasion), also using five-point Likert scores.

Using this validated pipeline, we generated the PASTA-Gen-30K dataset comprising 30 000 synthetic image–mask–text pairs (2000 per lesion type across 15 lesion categories), with consistent organ- and lesion-level annotations (appendix p20-22).

**Model development and statistical analysis**

PASTA was pretrained using a three-dimensional UNet encoder–decoder architecture combined with a multilayer perceptron classification head. The pretrained model was subsequently fine-tuned across multiple downstream oncologic imaging tasks, including non-contrast CT tumour identification, lesion segmentation under both full-data and few-shot settings, structured lesion attribute classification, tumour staging, survival prediction, and cross-modality transfer learning. Across all tasks, the pretrained encoder was retained and adapted using lightweight, task-specific output heads. Training was conducted using supervised objectives tailored to each task.

Model performance was assessed using task-appropriate metrics: area under the receiver operating characteristic curve (AUC) for tumour identification and survival prediction, Dice similarity coefficient (DSC) for segmentation, and accuracy and F1-score for structured lesion report generation. Five-fold cross-validation was applied throughout to ensure robust performance estimation, with repeated random subsampling employed for few-shot experiments. Statistical comparisons between models and experimental conditions were performed using paired analyses across cross-validation folds. Results are reported as mean values with variability across runs.

### Clinical decision support system development

PASTA-AID was developed as a workflow-integrated clinical decision support system based on task-specific models fine-tuned from the pretrained PASTA framework. System design was informed by close collaboration with radiologists to ensure clinical relevance and usability. The platform processes CT scans through an interactive interface to support three core functions: tumour screening, lesion segmentation, and structured radiology report generation. Two clinical workflows were implemented: (1) a screening-aid workflow for rapid tumour detection on non-contrast CT scans, and (2) a diagnosis-aid workflow for detailed lesion delineation and structured reporting on contrast-enhanced CT scans. Radiologists were able to review, edit, and refine both visual outputs (eg, segmentation masks) and generated report elements. The system was subsequently evaluated in a simulated clinical reader study to assess efficiency, diagnostic performance, and inter-observer agreement under clinically realistic reading conditions.

### Retrospective simulated clinical study design and statistical analysis

This retrospective clinical study was approved and registered in the Chinese Clinical Trial Registry (https://www.chictr.org.cn/) under the registration number ChiCTR2500101081. A simulated reader study was conducted to evaluate the clinical efficiency of PASTA-AID across three use cases: non-contrast CT tumour identification, lesion segmentation, and structured lesion report generation. All cases were drawn from the institutional cohorts described above and included cases of liver, pancreatic, and kidney cancers with matched control cases. Two junior and two senior radiologists participated in all experiments. Reader-study cases were not used for model fine-tuning.

For tumour identification, 60 tumour cases and 300 control scans were included per cancer type. In each study condition (solo and PASTA-AID–assisted), each radiologist reviewed 30 tumour cases and 150 control scans under both solo and PASTA-AID–assisted settings, with no case overlap between conditions. Radiologists completed each session within a fixed 30-minute time window. Performance was assessed using recall, precision, and case throughput. For lesion segmentation and structured report generation, 200 patients per cancer type were included, with cases evenly distributed among radiologists in both solo and assisted settings. In the assisted setting, radiologists refined model-generated lesion masks or structured reports, whereas in the solo setting, tasks were completed manually. Clinical efficiency was evaluated using task completion time. Agreement and reporting accuracy were quantified using the Dice similarity coefficient, per-attribute accuracy, and F1-score, where appropriate.

**Results**

We developed an integrated synthetic-to-clinic framework linking controllable lesion generation (PASTA-Gen), large-scale synthetic dataset construction (PASTA-Gen-30K), supervised pretraining of a tumour-centric foundation model (PASTA), and its clinical translation via a decision-support system (PASTA-AID) (figure 2, appendix p3-17). Guided by radiology-derived structured attributes, PASTA-Gen synthesised realistic malignant and benign lesions within anatomically healthy organs, producing paired image–mask–text data suitable for supervised representation learning. Using this framework, we constructed the publicly available PASTA-Gen-30K dataset comprising 30 000 synthetic 3D CT cases spanning ten malignant and five benign lesion categories, each with pixel-level lesion and organ annotations.

Leveraging this large-scale synthetic resource, PASTA was pretrained using a two-stage strategy involving lesion segmentation followed by vision–language alignment, enabling the model to learn tumour-centric representations that capture fine-grained lesion characteristics while preserving anatomical context.

To assess real-world applicability, we further translated PASTA into a workflow-integrated clinical decision-support system (PASTA-AID). The system supports rapid tumour detection on non-contrast CT and automated lesion segmentation and structured reporting on contrast-enhanced CT. Integrated into routine reading workflows,

PASTA-AID was designed to enhance diagnostic efficiency and consistency, compensate for experience-related performance gaps, and facilitate scalable clinical deployment across screening and diagnostic settings.

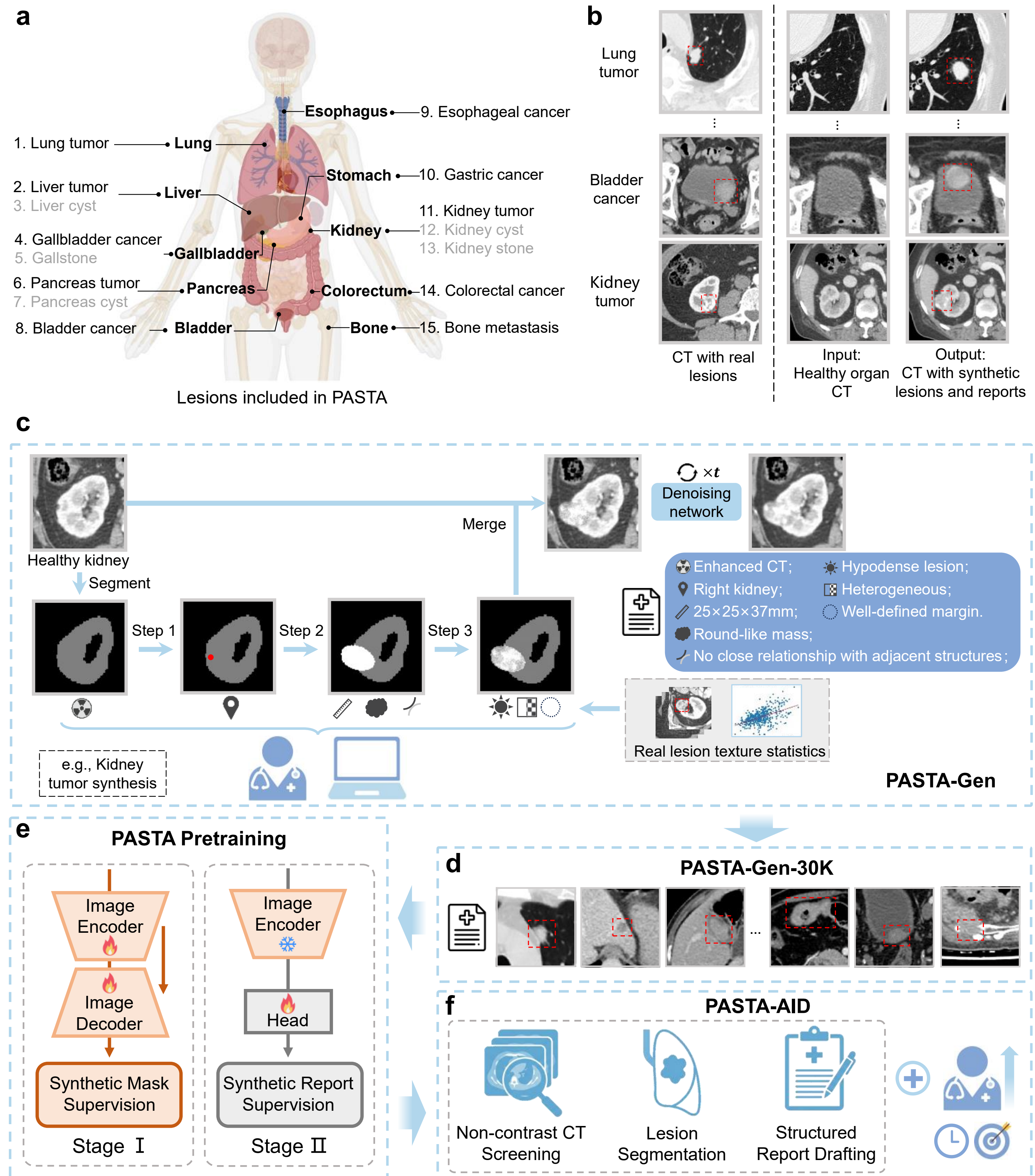

**Figure 2. Workflow of PASTA Model Development and Training Pipeline.**
**a**, Overview of organs and lesion types involved in PASTA training. **b**, Examples of lesions generated by PASTA-Gen from healthy organs. **c,** Lesion generation pipeline of PASTA-Gen. **d**, PASTA-Gen generates 30,000 masks and reports paired pan-tumour datasets, PASTA-Gen-30K. **e**, Two-stage pretraining of PASTA using the PASTA-Gen-30K dataset. **f,** Intergrating fine-tuned PASTA into PASTA-AID improves the efficiency and accuracy of radiologists across oncology tasks.

To characterise both algorithmic performance and clinical applicability, we conducted a staged evaluation of PASTA, combining benchmark testing across multiple oncologic tasks with workflow-integrated reader studies (figure 3). The model was first assessed for generalisability across downstream applications and subsequently deployed within the PASTA-AID decision-support system for retrospective clinical simulation analysis in two representative scenarios: non-contrast CT screening and contrast-enhanced CT diagnosis workflows.

In non-contrast CT tumour identification, PASTA was evaluated on liver, pancreatic, and kidney cancer cohorts collected at CMU, comprising 3864 scans in total. Using five-fold cross-validation, PASTA consistently achieved high discrimination performance across all three cancers, with AUC values ranging from 0.964 to 0.987, outperforming established pretrained models (Models Genesis[20] and SuPreM[21]) as well as a non-pretrained PASTA variant. The greatest improvement was observed in liver tumour detection, where PASTA improved AUC by 0.095 compared with the next-best model (figure 3b, appendix p29).

To examine clinical impact, we embedded PASTA into the PASTA-AID system and conducted a simulated reader study involving two junior and two senior radiologists. Under fixed 30-minute reading sessions, PASTA-AID significantly increased screening throughput across all cancer types, with case throughput increasing by 25.1% for liver, 11.1% for pancreatic, and 18.7% for kidney examinations compared with unaided reading.

Diagnostic performance also improved markedly. Pancreatic tumour recall increased from 61.3% to 92.7%, with the F1 score rising from 67.3% to 91.3%. For kidney tumours, precision improved from 81.2% to 96.2% and recall from 68.2% to 90.8%, yielding an F1 score of 93.2%. Liver tumour detection showed similar gains, with precision increasing from 71.3% to 96.2% and recall from 73.5% to 90.5%. Averaged across tumour types and readers, recall, precision, and F1 score increased by 23.7%, 16.8%, and 21.4%, respectively (figure 3c, appendix

p30).

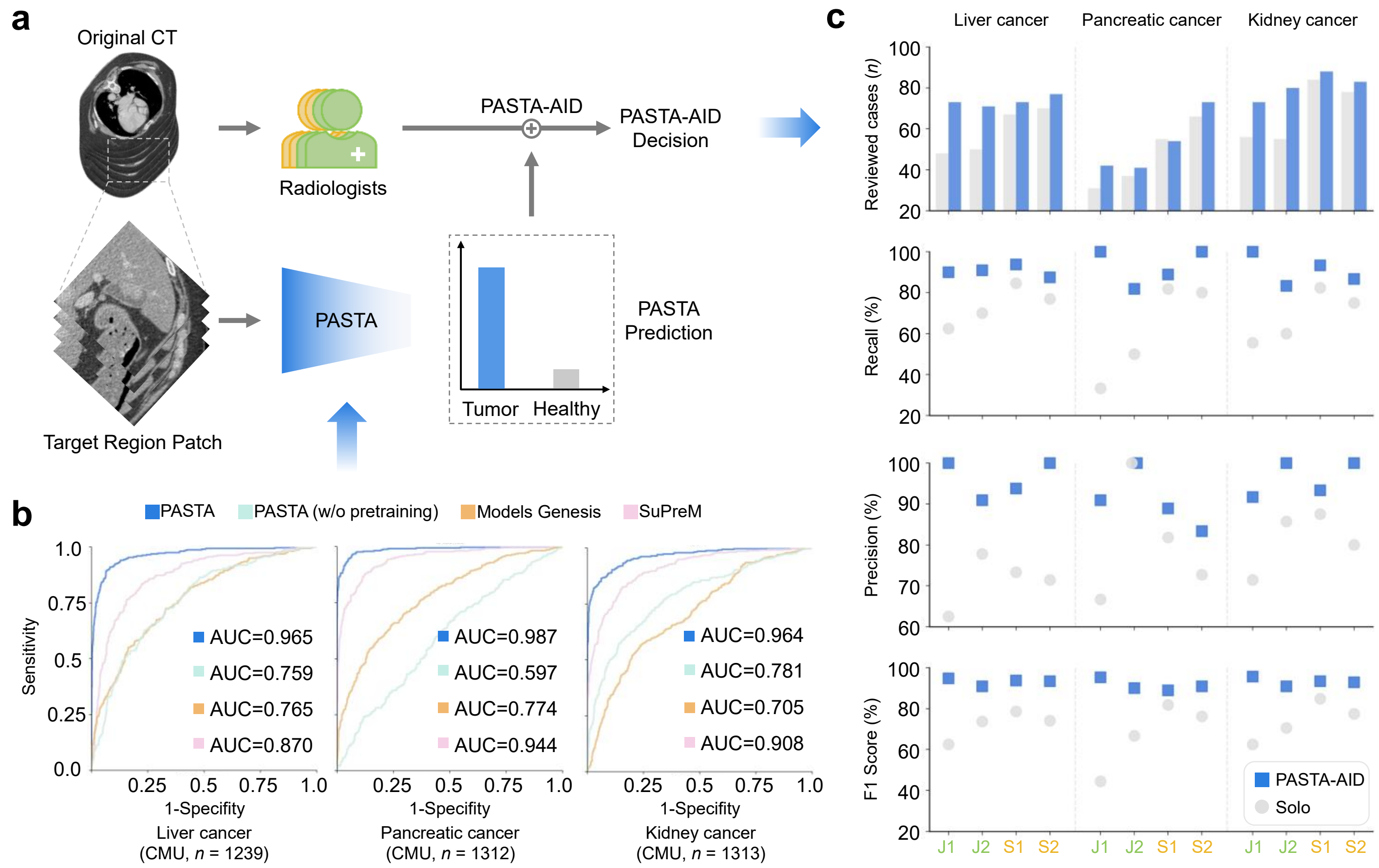

**Figure 3. Performance in Non-contrast CT Tumour Identification.**
Integration of PASTA fine-tuned on the non-contrast CT screening dataset into a clinical decision-support system, and evaluation of radiologists' efficiency and accuracy in tumour identification. **a**, Workflow of the PASTA-AID–assisted non-contrast CT tumour identification. **b**, Performance of different models in automatic non-contrast CT tumour screening, measured by AUC. **c**, Results of simulated high-workload non-contrast CT tumour screening within a fixed 30-minute reading window, with and without PASTA-AID assistance. J1 and J2 denote the two junior radiologists, and S1 and S2 denote the two senior radiologists.

We next evaluated PASTA for full-data lesion segmentation using a large cohort assembled from multiple centres and institutional sources comprising 1535 CT scans spanning 15 lesion types. Public datasets with high-quality masks were prioritised, including cohorts of lung, liver, pancreatic, colon, and kidney lesions from the Medical Segmentation Decathlon[22] and KiTS datasets[23], supplemented with manually annotated CMU cases for tumour types lacking public labels (eg, gallbladder, oesophageal, gastric, bladder cancers, and bone metastases), as well as representative benign lesions (figure 4).

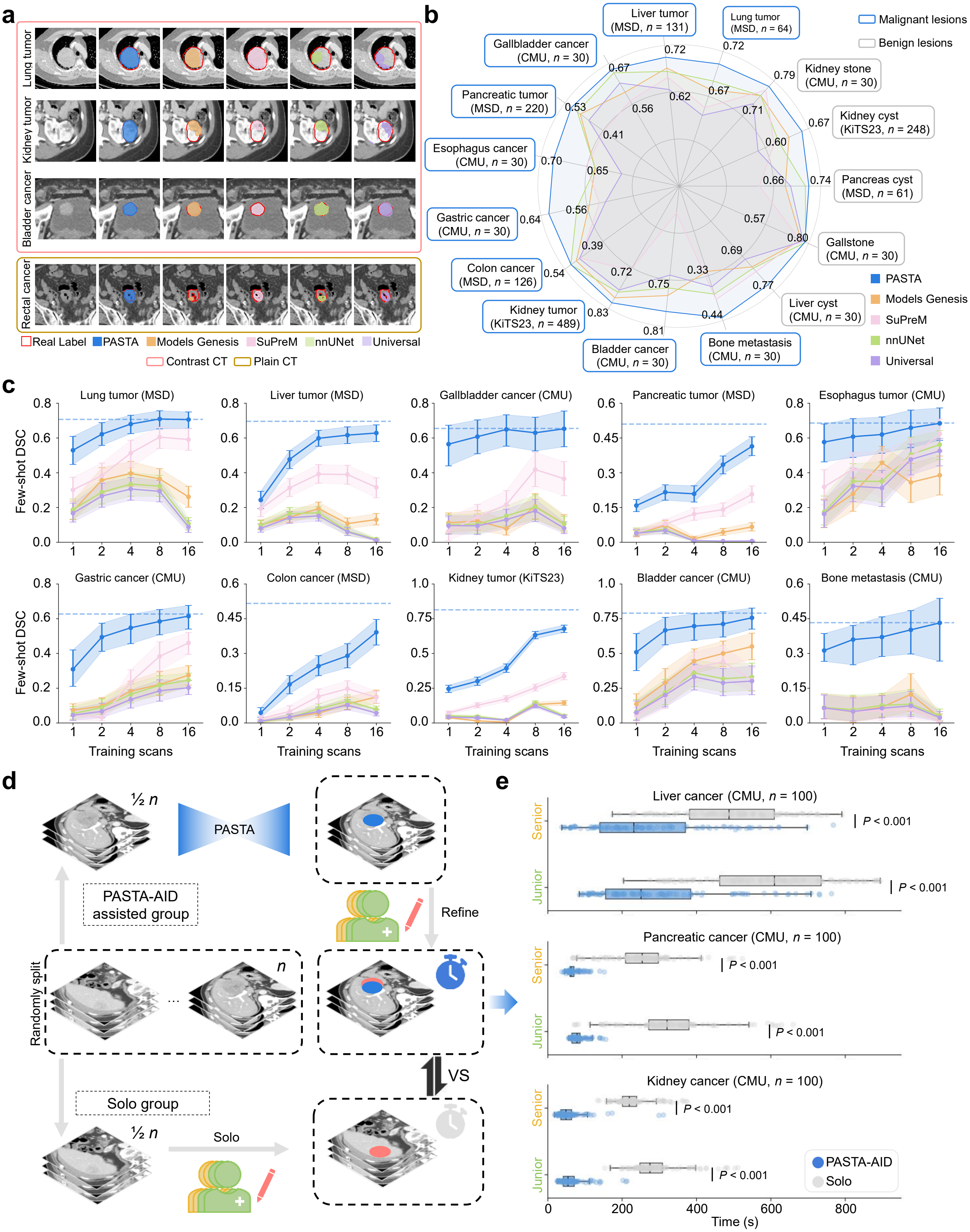

**Figure 4. Comparison on Lesion Segmentation.**
**a**, Example images of lesion segmentation results from various models. **b**, Comparison of model performance in lesion segmentation with sufficient data, measured by DSC. **c**, Lesion segmentation performance of models under few-shot settings, with blue dashed lines indicating PASTA's full-data training results. Error bands denote 95% confidence intervals. **d**, Lesion segmentation reader study design. Radiologists reviewed randomly assigned scans under PASTA-AID–assisted or unassisted conditions. **e**, Lesion mask annotation time of radiologists with different levels of experience, with and without PASTA-AID assistance. Box plots show the median (center line), interquartile range (IQR, box limits represent 25th and 75th percentiles), and whiskers extending to 1.5 × IQR; outliers are shown as individual points. Annotation time differences between groups were assessed using a linear mixed-effects model with doctor as a random intercept, and corresponding p-values are reported.

PASTA was benchmarked against established segmentation frameworks (nnUNet[26] and Universal[27]) and pretrained foundation models (Models Genesis[20] and SuPreM[21]). Across all 15 segmentation tasks, PASTA consistently outperformed other pretrained models, achieving Dice similarity coefficients (DSC) ranging from 0.433 to 0.814, except for gallstone segmentation where nnUNet performed marginally better (appendix p31-36). Significant improvements over the next-best model were observed for seven tumour types, including lung (+1.9%), liver (+2.3%), pancreatic (+1.9%), oesophageal (+3.7%), gastric (+4.6%), kidney (+1.4%), and bone metastases (+4.4%). Notably, the largest gains occurred in tumours traditionally associated with poor segmentation performance, such as gastric cancer and bone metastases.

To assess label efficiency, we further evaluated PASTA under few-shot conditions using the same datasets. With extremely limited training data ($K \in \{1,2,4,8,16\}$) and only 2000 training iterations, PASTA consistently outperformed all baselines, with absolute DSC improvements ranging from 0.025 to 0.463 (appendix p31-36). Under ultra-low data regimes ($n \leq 2$), PASTA maintained strong performance. For gallbladder cancer (n=2), PASTA achieved a DSC of 0.608, significantly outperforming SuPreM and approaching the performance achieved under full-data training. Similarly, for bladder cancer (n=2), PASTA reached a DSC of 0.667, exceeding Models Genesis by 37.7% and nearing the full-data benchmark.

We further translated these segmentation capabilities into clinical practice through PASTA-AID and evaluated workflow efficiency in a retrospective simulated diagnosis-aid study. Two junior and two senior radiologists segmented contrast-enhanced CT scans for liver, pancreatic, and kidney cancers with and without system assistance. PASTA-AID markedly reduced per-case segmentation time across all cancers, with reductions of up to 74.5% for senior radiologists and 78.2% for junior radiologists (appendix p38). In addition to efficiency gains, PASTA-AID significantly improved segmentation quality for less-experienced readers. Compared with gold-standard annotations from senior radiologists, junior readers achieved 11.8–12.2% higher Dice similarity when assisted by the system (appendix p39).

We further evaluated PASTA for structured lesion report generation using 1535 annotated scans spanning 15 lesion types, following the standardised reporting schema defined by PASTA-Gen. The task was formulated as a multi-class classification task for each lesion attribute. PASTA consistently outperformed competing foundation models across all attributes and lesion types, achieving higher accuracy and F1 scores than Models Genesis and SuPreM, except for the invasion attribute where performance was comparable. Compared with the next-best model, PASTA showed significant improvements in both accuracy and F1 score for key attributes, including lesion shape, density, heterogeneity, and surface characteristics (figure 5, appendix p40).

When integrated into the PASTA-AID system, these capabilities translated into measurable clinical efficiency gains. In the retrospective simulated reader study, PASTA-AID automatically generated draft structured reports that radiologists could review and refine. Across tumour types and reader experience levels, report preparation time was reduced by 15.7–36.5%. Among junior radiologists, system assistance also improved concordance with senior readers' reports, although this difference did not reach statistical significance.

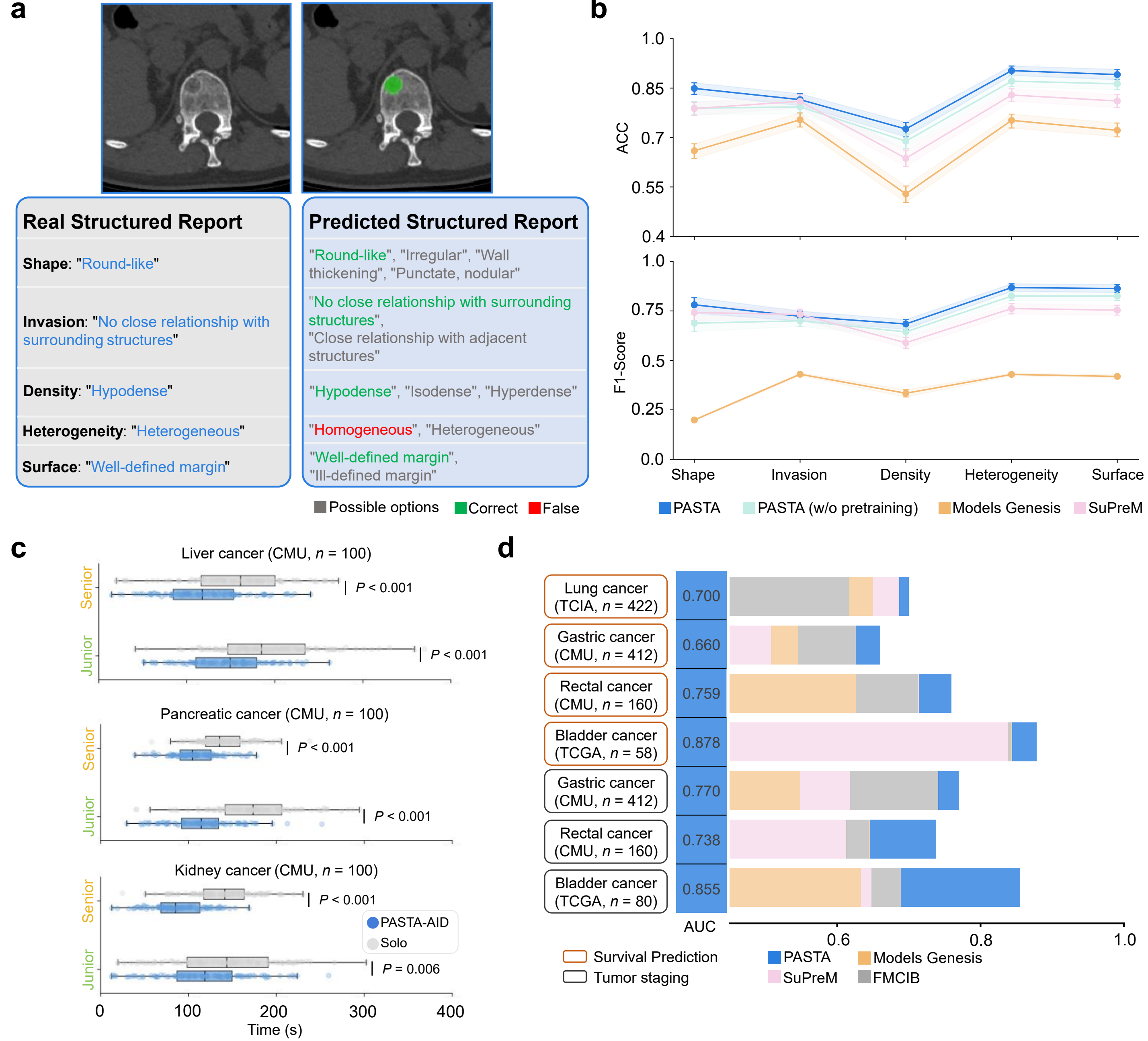

**Figure 5. Evaluation of Structured Lesion Report Generation.**
**a**, Example of real and predicted structured lesion reports for bone metastasis generated by PASTA. **b**, Comparison of Accuracy (ACC) and F1-scores for five structured report attributes across different models. Error bands denote 95% confidence intervals. **c.** Reporting time of radiologists with different levels of experience, under conditions with and without PASTA-AID assistance. Box plots show the median (center line), interquartile range (IQR, box limits represent 25th and 75th percentiles), and whiskers extending to 1.5 × IQR; outliers are shown as individual points. Report time differences between groups were assessed using a linear mixed-effects model with doctor as a random intercept, and corresponding p-values are reported. **d**, Performance of models in tumour staging and survival prediction across various tumour types.

Beyond descriptive tasks, we evaluated PASTA for tumour staging and survival prediction (figure 5d,

appendix p41). Staging performance was assessed for gastric, rectal, and bladder cancers using public and institutional datasets, with PASTA achieving AUC values ranging from 0.738 to 0.855 (gastric cancer: stage I-II vs. stage III-IV; rectal cancer: stage I-III vs. stage IV; bladder cancer: stage I-II vs. stage III-IV). Significant improvements over the next-best model were observed across all tumour types, including gains of 0.029 for gastric cancer, 0.092 for rectal cancer, and 0.166 for bladder cancer. For survival prediction, PASTA demonstrated strong discriminatory ability across lung, gastric, rectal, and bladder cancers, achieving AUCs between 0.660 and 0.878 and outperforming comparison models (lung tumor: overall survival (OS) < 2 yrs vs. OS ≥ 2 yrs; gastric cancer: OS < 2 yrs vs. OS ≥ 2 yrs; rectal cancer: OS< 3 yrs vs. OS ≥ 3yrs; bladder cancer: OS < 3 yrs vs. OS ≥ 3 yrs), including FMCIB[28], particularly for gastric and rectal cancers. These findings show that representations learned from synthetic data extend beyond image-level tasks to clinically relevant prognostic modelling.

We further examined PASTA's ability to transfer across imaging modalities by evaluating performance on MRI brain and liver tumour datasets under limited data conditions. Despite being pretrained exclusively on CT data, PASTA consistently outperformed other models in few-shot MRI segmentation (appendix p36-37). On cross-domain brain MRI data, the model achieved a DSC of 0.504 at n=16, significantly exceeding SuPreM. Similarly, on MRI liver tumour segmentation, PASTA reached a DSC of 0.603, outperforming SuPreM by 13.4%. These results highlight PASTA's strong cross-modality transfer capability, suggesting that the model learns generalisable representations that effectively distinguish normal and abnormal tissue patterns across diverse imaging modalities.

Before constructing the large-scale PASTA-Gen-30K dataset, we conducted a systematic quality assessment to verify the realism and text–image consistency of PASTA-Gen outputs (appendix p19). Four radiologists independently evaluated synthetic and real lesions in a blinded setting. Across all lesion types, synthetic images achieved high realism scores, with mean ratings ranging from 4.23 to 4.73 on a five-point scale, indicating that most generated samples were perceived as highly realistic. Notably, for kidney cysts, synthetic images slightly outperformed real data in perceived realism.

We further examined the consistency between generated images and their paired structured reports. Radiologists assessed alignment across five clinically relevant attributes, including lesion shape, density, heterogeneity, surface characteristics, and anatomical adjacency. Across all 15 lesion types, mean consistency scores ranged from 4.54 to 4.93, reflecting excellent agreement between visual features and textual descriptions. Particularly high consistency was observed for cystic and calcified lesions, while tumour-like lesions also exceeded the threshold for high accuracy.

This systematic validation confirmed that PASTA-Gen produces synthetic CT images and structured reports that closely resemble real-world radiological characteristics, thereby providing a reliable foundation for subsequent large-scale dataset construction and downstream model development.

## Discussion

This study presents PASTA, a synthetic data-enabled artificial intelligence system developed to support pan-tumour CT screening and diagnosis across multiple cancer types. A central innovation is PASTA-Gen, a generative framework capable of synthesising realistic malignant and benign lesions across ten organs, which enabled the construction of the large-scale PASTA-Gen-30K synthetic dataset with pixel-level lesion masks and structured reports.

Using a two-stage pretraining strategy based on lesion segmentation and vision–language alignment, PASTA effectively leveraged synthetic data to overcome long-standing challenges related to annotation scarcity and data privacy. Across diverse imaging datasets and modalities, including MRI, the model demonstrated robust generalisability and supported a wide range of clinically relevant tasks, including non-contrast CT tumour detection, lesion segmentation, tumour staging, survival prediction, and structured reporting.

Beyond algorithmic evaluation, we translated PASTA into a workflow-aligned clinical decision-support system (PASTA-AID) and assessed its performance in a retrospective simulated clinical study. The system significantly improved diagnostic efficiency, reporting workflow, and agreement between junior and senior radiologists, highlighting its practical utility in routine clinical practice. These findings suggest that synthetic data-trained AI systems can meaningfully support radiologists under high-workload conditions and have the potential to enhance diagnostic accuracy and clinical decision-making.

Compared with existing AI imaging models that are typically limited to specific tumour types or rely on large-scale real-world annotations, PASTA offers a scalable pan-tumour solution enabled by synthetic data. By covering a broad spectrum of organs and lesion types, this framework addresses the fundamental data scarcity challenge faced in rare cancers and supports more equitable development and deployment of AI across oncology domains.

First, by generating large volumes of publicly available and precisely controlled image–text pairs, PASTA-Gen offers a practical solution to the long-standing shortage of real-world datasets with pixel-level lesion annotations and comprehensive radiology reports. Manual annotation is time-consuming and resource-intensive, and data sharing is further constrained by privacy regulations. High-quality synthetic data therefore provides a complementary pathway for enabling methodological development and clinical translation without compromising patient confidentiality.

Second, large-scale imaging foundation models spanning multiple organs and tumour types remain rare, primarily due to limited access to annotated data. However, shared biological characteristics across malignancies suggest that unified modelling strategies may yield clinically meaningful insights, as demonstrated in pathology and molecular oncology research. By systematically simulating tumours across ten organs, PASTA represents a step towards truly pan-tumour 3D imaging foundation modelling, overcoming the fragmentation of real-world datasets.

Importantly, the pan-tumour design of PASTA translates into tangible clinical benefits across multiple downstream tasks. In few-shot learning settings—highly relevant for rare cancers and resource-limited institutions—PASTA maintained high segmentation accuracy with only one or two labelled cases, substantially reducing data collection and annotation requirements. Beyond segmentation, the model also supported structured reporting, tumour staging, and survival prediction.

When deployed within the PASTA-AID decision-support system, these capabilities translated into measurable workflow improvements without disrupting routine reading practice. In simulated non-contrast CT screening under time-limited, high-workload conditions, PASTA-AID increased screening throughput, recall, and precision. In diagnostic workflows, it markedly reduced segmentation time and reporting effort while improving agreement between junior and senior radiologists.

Notably, these findings underscore the potential of AI-assisted systems to support less-experienced clinicians and reduce disparities between tertiary centres and resource-constrained institutions. By narrowing expertise gaps and alleviating diagnostic burden, PASTA-AID may contribute to more equitable access to high-quality oncological imaging care. Collectively, these results position PASTA as a scalable and adaptable framework for clinically oriented AI development.

Despite these promising outcomes, several limitations should be acknowledged. First, although the synthetic data generation pipeline was rigorously validated by radiologists, simulated lesions may still differ subtly from real-world disease complexity. Second, the model's performance in rare tumour subtypes and atypical lesion presentations warrants further validation using clinically acquired datasets. Finally, although we incorporated multiphase CT data from multiple institutions, inclusion of more diverse patient populations would further strengthen model robustness and mitigate potential scanner- or population-related biases.

In conclusion, the release of PASTA and PASTA-Gen-30K represents an important advance in the development of pan-tumour imaging foundation models, addressing critical challenges related to data scarcity, privacy, and generalisability. By leveraging synthetic training data, PASTA enhances tumour screening and lesion annotation performance while supporting a wide range of downstream clinical tasks. Through the PASTA-AID system, we further provide direct evidence of translational potential, demonstrating meaningful improvements in efficiency, diagnostic accuracy, and clinician support. Future work will focus on integrating additional imaging modalities, refining synthetic data generation to better capture complex clinical variability, and conducting large-scale, multi-centre real-world validation studies. As AI-driven medical imaging continues to evolve, this work lays the groundwork for more generalisable, data-efficient, and clinically impactful models, particularly for settings where access to large annotated datasets remains limited.

**Contributors** Conceptualization: W.L., H.C., X.Z., P.R., S.Z., and Z.W. Methodology: W.L., H.C., Z.Z., and L.L. Investigation: W.L., H.C., Z.Z., L.L., Q.X., Y.G., P.G., Y.J., C.W., G.W., T.X., and Y.Z. Visualization: W.L. and H.C. Funding acquisition: H.C., X.Z., S.Z., and Z.W. Supervision: X.Z., S.Z., and Z.W. Writing: W.L., H.C., and L.L.

**Data Sharing**

To facilitate reproducibility and further research, the PASTA-Gen-30K synthetic dataset, comprising 30 000 paired 3D CT images, lesion masks, and structured reports, is publicly available at https://huggingface.co/datasets/LWHYC/PASTA-Gen-30K. All publicly available datasets used in this study can be accessed through their original sources, including the Medical Segmentation Decathlon (http://medicaldecathlon.com/), KiTS23 (https://kits-challenge.org/kits23/), Lung1

(https://www.cancerimagingarchive.net/collection/nsclc-radiomics/), TCGA-BLCA (https://www.cancerimagingarchive.net/collection/tcga-blca/), and ATLAS (https://atlas-challenge.u-bourgogne.fr/atlas/). Structured report annotations for public datasets and lesion centre-point annotations for Lung1 and TCGA-BLCA are available at https://github.com/LWHYC/PASTA. The complete PASTA model, pretrained weights, and full training and evaluation pipeline are also publicly accessible at https://github.com/LWHYC/PASTA.

Additional institutional imaging data and associated clinical records were obtained from The First Hospital of China Medical University. Owing to privacy and ethical restrictions, these data cannot be publicly shared. This retrospective clinical study was reviewed and registered in the Chinese Clinical Trial Registry (ChiCTR2500101081).

**Declaration of interests**

The authors declare no competing interests.

**Acknowledgements** This work was supported by the National Natural Science Foundation of China (62301311) (X.Z.), Shanghai Municipal Commission of Economy and Informatization (204694) (S.Z.), the Noncommunicable Chronic Diseases ‒ National Science and Technology Major Project (2023ZD0501500) (Z.W.), the National Natural Science Foundation of China (U123A20457) (Z.W.), National Natural Science Foundation of China (82203199) (H.C.).

## Reference

1. Bray F, Laversanne M, Sung H, et al. Global cancer statistics 2022: GLOBOCAN estimates of incidence and mortality worldwide for 36 cancers in 185 countries. *CA: a cancer journal for clinicians* 2024; **74**(3): 229-63.
2. Kleeff J, Ronellenfitsch U. AI and imaging-based cancer screening: getting ready for prime time. *Nature medicine* 2023; **29**(12): 3002-3.
3. Cao K, Xia Y, Yao J, et al. Large-scale pancreatic cancer detection via non-contrast CT and deep learning. *Nature medicine* 2023; **29**(12): 3033-43.
4. Ardila D, Kiraly AP, Bharadwaj S, et al. End-to-end lung cancer screening with three-dimensional deep learning on low-dose chest computed tomography. *Nature medicine* 2019; **25**(6): 954-61.
5. McKinney SM, Sieniek M, Godbole V, et al. International evaluation of an AI system for breast cancer screening. *Nature* 2020; **577**(7788): 89-94.
6. Preetha CJ, Meredig H, Brugnara G, et al. Deep-learning-based synthesis of post-contrast T1-weighted MRI for tumour response assessment in neuro-oncology: a multicentre, retrospective cohort study. *The Lancet Digital Health* 2021; **3**(12): e784-e94.
7. Lipkova J, Chen RJ, Chen B, et al. Artificial intelligence for multimodal data integration in oncology. *Cancer cell* 2022; **40**(10): 1095-110.
8. Moor M, Banerjee O, Abad ZSH, et al. Foundation models for generalist medical artificial intelligence. *Nature* 2023; **616**(7956): 259-65.
9. Zhang S, Metaxas D. On the challenges and perspectives of foundation models for medical image analysis. *Medical image analysis* 2024; **91**: 102996.
10. Brown T, Mann B, Ryder N, et al. Language models are few-shot learners. *Advances in neural information processing systems* 2020; **33**: 1877-901.
11. Acosta JN, Falcone GJ, Rajpurkar P, Topol EJ. Multimodal biomedical AI. *Nature Medicine* 2022; **28**(9): 1773-84.
12. Rajpurkar P, Topol EJ. A clinical certification pathway for generalist medical AI systems. *The Lancet* 2025; **405**(10472): 20.
13. Chen RJ, Ding T, Lu MY, et al. Towards a general-purpose foundation model for computational pathology. *Nature Medicine* 2024; **30**(3): 850-62.
14. Wang X, Zhao J, Marostica E, et al. A pathology foundation model for cancer diagnosis and prognosis prediction. *Nature* 2024; **634**(8035): 970-8.
15. Hua S, Yan F, Shen T, Ma L, Zhang X. Pathoduet: Foundation models for pathological slide analysis of H&E and IHC stains. *Medical Image Analysis* 2024; **97**: 103289.
16. Lei W, Xu W, Li K, Zhang X, Zhang S. MedLSAM: Localize and segment anything model for 3D CT images. *Medical Image Analysis* 2025; **99**: 103370.
17. Wang G, Wu J, Luo X, Liu X, Li K, Zhang S. Mis-fm: 3d medical image segmentation using foundation models pretrained on a large-scale unannotated dataset. *arXiv preprint arXiv:230616925* 2023.
18. Wald T, Ulrich C, Lukyanenko S, et al. Revisiting MAE pre-training for 3D medical image segmentation. *arXiv preprint arXiv:241023132* 2024.
19. Jiang Y, Sun M, Guo H, et al. Anatomical invariance modeling and semantic alignment for self-supervised learning in 3d medical image analysis. Proceedings of the IEEE/CVF International Conference on Computer Vision; 2023; 2023. p. 15859-69.
20. Zhou Z, Sodha V, Pang J, Gotway MB, Liang J. Models genesis. *Medical image analysis* 2021; **67**: 101840.

21. Li W, Yuille A, Zhou Z. How well do supervised 3d models transfer to medical imaging tasks? *arXiv preprint arXiv:250111253* 2025.
22. Antonelli M, Reinke A, Bakas S, et al. The medical segmentation decathlon. *Nature communications* 2022; **13**(1): 4128.
23. Heller N, Isensee F, Maier-Hein KH, et al. The state of the art in kidney and kidney tumor segmentation in contrast-enhanced CT imaging: Results of the KiTS19 challenge. *Medical image analysis* 2021; **67**: 101821.
24. Aerts HJ, Velazquez ER, Leijenaar RT, et al. Decoding tumour phenotype by noninvasive imaging using a quantitative radiomics approach. *Nature communications* 2014; **5**(1): 4006.
25. Kirk S, Lee Y, Lucchesi F, et al. The Cancer Genome Atlas Urothelial Bladder Carcinoma Collection (TCGA-BLCA)(Version 8)[Data set]. *The Cancer Imaging Archive* 2016; **10**: K9.
26. Isensee F, Jaeger PF, Kohl SA, Petersen J, Maier-Hein KH. nnU-Net: a self-configuring method for deep learning-based biomedical image segmentation. *Nature methods* 2021; **18**(2): 203-11.
27. Liu J, Zhang Y, Chen J-N, et al. Clip-driven universal model for organ segmentation and tumor detection. Proceedings of the IEEE/CVF International Conference on Computer Vision; 2023; 2023. p. 21152-64.
28. Pai S, Bontempi D, Hadzic I, et al. Foundation model for cancer imaging biomarkers. *Nature machine intelligence* 2024; **6**(3): 354-67.

# Supplementary appendix

**eMethods**

## 1. Development of PASTA-Gen

PASTA-Gen is a lesion editing model designed to simulate lesions on scans of target organs with healthy anatomy. To develop PASTA-Gen, we first constructed a universal structured lesion report template, followed by the collection and annotation of real lesion data as a reference set. Using this reference data and the structured template, we iteratively refined the model through close collaboration with radiologists, ensuring its clinical relevance and accuracy. Finally, a denoising network was incorporated to enhance the realism of the generated images.

### 1.1 Structured Lesion Report Template

Building on previous studies on structured radiology reporting[1], we synthesized insights from existing research and analyzed a large collection of in-house radiology reports across various disease types. While no universal standard currently exists, radiologists consolidated lesion descriptions for solid tumours into eight key attributes, including enhancement status, location, size, shape, density, heterogeneity, surface, and invasion. These attributes form a comprehensive and objective template for characterizing lesions, ensuring consistency and enabling generalization across diverse lesion types (eTable 1).

One of these attributes, "density," typically relies on assessments from multiple CT modalities. For instance, in contrast-enhanced CT scans, radiologists often integrate information from different imaging phases to provide a nuanced evaluation of a lesion’s density. However, as our study focuses on developing a cross-phase CT analysis model, the concept of "density" in PASTA-Gen refers to a relative assessment within a single-phase CT image. Specifically, it captures the density of the lesion area relative to the surrounding normal tissues in the given phase. This approach enhances the model’s adaptability in representing lesions with complex or variable densities.

### 1.2 Real Lesion Reference Data

To ensure comprehensive modeling and evaluation of synthetic lesion data, we developed a diverse reference dataset encompassing 15 target lesion types across 10 organs. This dataset includes lesion segmentations and structured reports derived from both public sources and in-house data from the First Hospital of CMU (eTable 4). For instances with incomplete labels, two senior radiologists meticulously supplemented the missing information, ensuring the dataset's accuracy and clinical relevance. We reformat all CT scans so that the first axis points from right to left, the second from anterior to posterior, and the third from inferior to superior. We then resample the spacing in all directions to 1 mm using linear spline interpolation.

### 1.3 Modeling Process for PASTA-Gen

The development of PASTA-Gen involved a multidimensional statistical analysis of collected real reference lesions, combined with iterative feedback from experienced radiologists. This collaborative approach enabled us to systematically model the eight structured attributes outlined in our framework.

Lesion size distributions were statistically represented using a log-normal distribution across various axes, with random sampling used to generate realistic dimensions. Lesion shapes were modeled based on the specific morphological characteristics of each type, such as the round-like shape of liver cysts or the wall thickening observed in gastric cancer. These shape parameters ensure the model captures the clinical nuances of different lesions. Density attributes were linearly modeled with controlled variability, determined by calculating the intensity of normal tissues surrounding the lesion on target organ scans. The lesion density was then derived by sampling from these calculated differences, ensuring alignment with clinical observations. Heterogeneity attributes, including brightness deviation and texture features, were modeled based on radiologist input. For instance, heterogeneous density patterns were simulated to reflect clinical variability. For solid tumours in parenchymal organs like liver and kidney tumours, we emphasized the boundary between the tumour and surrounding tissues, controlling this characteristic by adjusting the degree of Gaussian blur applied to the tumour edges. For hollow organs such as those in the gastrointestinal or urinary systems, surface features were modeled to highlight serosal roughness and outer surface irregularities. Invasion was modeled based on whether the lesion extended beyond the target organ into adjacent tissues. For benign lesions, invasion into surrounding organs was explicitly avoided to maintain consistency with clinical presentations. Each simulated lesion is accompanied by a corresponding structured report containing eight attributes, with each attribute reflecting one of its possible values (eTable 1). This comprehensive approach ensures that PASTA-Gen generates synthetic lesions with high fidelity and clinical relevance, capable of capturing the diverse characteristics of real-world lesions.

### 1.4 Collection and Preprocessing of Template CT Scans and Radiology Report

To ensure the diversity and scalability of PASTA-Gen, we collected an additional 10,767 in-house CT scans to serve as the template scan set, which include both contrast-enhanced and plain phases, covering various parts of the body, along with their corresponding radiology reports (eTable 2). We reformat all CT scans so that the first axis points from right to left, the second from anterior to posterior, and the third from inferior to superior. We then resample the spacing in all directions to 1 mm using linear spline interpolation. We use TotalSegmentator[2] to segment the standardized scans. This tool segments 104 classes of organs, encompassing all the organs for which PASTA-Gen simulates lesions.

For the template CT scans. Based on the presence of specific organs, we only retain the following three study types of CT scans: (1) Chorax CT: must include the T1 vertebra, left upper lobe of the lung, and right upper lobe of the lung, but not include the L5 vertebra. (2) Abdomen-pelvis CT: must include the T8 vertebra and the bladder but not include the T1 vertebra. (3) Thorax-abdomen-pelvis CT: must include the T1 vertebra, left upper lobe of the lung, and right upper lobe of the lung and the bladder. Next, we determine whether each selected CT scan is contrast-enhanced by evaluating the Hounsfield unit (HU) values of the aorta and inferior vena cava. If the average HU of both the aorta and inferior vena cava is less than 80, the scan is classified as non-contrast. Otherwise, it is classified as contrast-enhanced. Based on the diagnostic modalities typically used by radiologists for different types of lesions, we selected the appropriate scanning regions and modalities for simulating each type of lesion (Supplementary Table 8).

To ensure that PASTA-Gen simulates target lesions on healthy organs, we implemented a filtering process based on the radiology report associated with each CT scan. For the target organ in each CT scan, the criteria for being classified as healthy are as follows: (1) the volume of the organ is larger than 4000 $mm^3$, and (2) the organ was not mentioned in the suspicious findings section of the radiology report, as determined by keyword search. By applying these criteria, we ensure that only healthy organs are used for simulating lesions (eTable 3).

### 1.5 Details of Denoising Network

The denoising network in PASTA-Gen aims to optimize the initially synthesized lesions. This network is based on a 3D medical denoising diffusion probabilistic model (DDPM)[3,4], which involves adding random Gaussian noise to the original image and learning how to remove it. We leverage this property to eliminate unnatural parts of the simulated lesions. During the training phase, the input image size is 128×128×128. We concatenate the corresponding organ segmentation mask as the input condition for the model (eFigure 1). We set the total timesteps $T$ to 1000 and utilize the $L_1$ loss to measure the difference between predicted and added noise. The model channel number is set to 16. We use an AdamW optimizer[5] with an initial learning rate of $1\times10^{-4}$ and weight decay of $1\times10^{-5}$. We employ a total batch size of 8 across 4 NVIDIA GTX4090 GPUs. During the lesion generation stage, the initially simulated lesions are input into the denoising network for $t = 5$ optimization iterations. A sliding window approach is applied to ensure comprehensive and seamless optimization across the entire image volume.

### 1.6 Systematic Evaluation of Synthetic Data Quality

The realism of generated images is a critical determinant of the utility of synthetic data. Accordingly, at the outset of the study, we conducted a systematic evaluation of both the image realism and text-image consistency of PASTA-Gen outputs (eTable 6). A panel of four radiologists assessed the realism of PASTA-Gen–generated images and the accuracy of their paired structured text descriptions (eFigure 2). For image realism, real and generated data for each lesion type were mixed in a 1:1 ratio, resulting in a combined dataset of 1,180 cases. All four radiologists performed a blind Turing test, rating each sample on a scale of 1 (completely unrealistic) to 5 (indistinguishable from real CT images).

**2. Dataset Curation for PASTA-Gen-30K**

Using PASTA-Gen, we simulated 2,000 instances for each of the 15 target lesion types, resulting in a dataset of 30,000 lesion image-mask-text pairs. Since the template scan set includes organ segmentation results generated by TotalSegmentator, the labels for the simulated images retain all 10 target organ labels alongside the lesion labels, totaling 25 categories (eTable 15).

**3. Details of PASTA Pretraining**

The PASTA model leverages a 3D UNet[6] architecture for its encoder and decoder, with a three-layer MLP serving as the classification head. The encoder and decoder of the 3D UNet architecture each consist of five convolutional blocks. Each convolutional block is composed of two sub-blocks, and within each sub-block, there is a 3D convolution operation with a kernel size of 3×3×3 and a stride of 1 along each dimension. Each convolution is followed by a 3D Instance Normalization[7] activation function. In the encoder, each convolutional block is followed by a max pooling operation with a kernel size of 2×2×2, while in the decoder, each block is followed by a transposed convolution operation with a stride of 2, which up-samples the spatial resolution. The MLP classification head consists of an input layer with 1024 channels, followed by two hidden layers with channel sizes of 512 and 256, and a final output layer mapping to the number of classes. Each intermediate layer is followed by a LeakyReLU activation function.

During the initial segmentation pretraining stage, we employed the nnUNet framework[8], which has demonstrated state-of-the-art performance across various biomedical image segmentation tasks[9,10]. The training target is to segment all 25 categories in PASTA-Gen-30K. The training process used an initial learning rate of $1\times10^{-2}$ and was conducted over 2,000 epochs, with each epoch comprising 250 iterations. Random cropping was applied to extract 3D patches of size 128×128×128 voxels for training. The loss function combined Cross-Entropy Loss and Dice Loss[11], optimizing for both pixel-wise classification accuracy and volumetric overlap. Training was conducted with a total batch size of 16, distributed across 8 NVIDIA GTX 4090 GPUs.

In the second stage of lesion attribute classification pretraining, the 3D UNet component is frozen, and only the MLP connected to the encoder is trained. The training objective is to predict lesion attributes from the structured reports in the PASTA-Gen-30K dataset, including Shape, Density, Heterogeneity, Surface, and Invasion attributes. Each attribute comprises fixed categories: Shape (4 classes), Density (3 classes), and Heterogeneity, Surface, and Invasion (2 classes each). The task is formulated as a multi-class classification problem. The initial learning rate is set to $1\times10^{-3}$, and the training runs for a total of 100,000 iterations. The loss function used is Cross-Entropy Loss. During training, patches of size 96×96×96 voxels, centered on the lesion, are cropped and fed into the network to focus on the lesion region. Training was conducted with a total batch size of 64, distributed across 4 NVIDIA GTX 4090 GPUs.

**4. Competing Methods and Baselines**

We compare PASTA to 5 comparison approaches. Models Genesis[12] was a released model pretrained on 623 Chest CT scans in LUNA 2016[13]. It utilized a 3D UNet architecture and employed a self-supervised learning approach by recovering the original sub-volumes of images from their transformed versions. SuPreM[14] was pretrained in a supervised segmentation setting on the AbdomenAtlas 1.1 dataset[15] which comprises 9,262 CT volumes with detailed annotations of 25 anatomical structures and pseudo annotations for seven tumour types.

SuPreM incorporates multiple backbone architectures, and we selected the 3D UNet model for evaluation as it demonstrated the best performance in SuPreM experimental results. FMCIB[16] used a Resnet[17] as the backbone and was pretrained with a comprehensive dataset of 11,467 radiographic lesions in DeepLesion[18]. It is tailored for cancer imaging biomarker discovery by contrasting volumes with and without lesions. nnUNet[23] and Universal[24] are both state-of-the-art frameworks for biomedical image segmentation.

## 5. Construction of PASTA-AID

To bridge algorithmic advances with clinical applicability, we developed PASTA-AID, a clinical decision support system built upon the PASTA foundation model. PASTA-AID was designed in close collaboration with radiologists to integrate seamlessly into existing workflows, aiming to reduce interpretation time and improve diagnostic consistency without replacing physicians' expertise.

### 5.1 System Design and Implementation

PASTA-AID was implemented in Python (v3.9.7) with an HTML-based interactive interface (built using Flask-based rendering, matplotlib v3.10.0, seaborn v0.13.2, hiddenlayer v0.3). The system ingests NIfTI-format CT scans and leverages PASTA for tumour screening, lesion segmentation, and report generation (implemented with torch v2.2.0, SimpleITK v2.4.1, nibabel v5.3.2, numpy v2.2.2, pandas v2.2.3, Scikit-learn v1.6.1, scipy v1.15.1). Utility packages such as tqdm v4.66.4 and Requests v2.32.3 were used for progress monitoring and data handling. The interface displays three orthogonal CT views (axial, coronal, sagittal) and allows users to interactively scroll through slices. In the non-contrast CT screening setting, PASTA-AID outputs the estimated cancer probability for the target organ in addition to visual overlays. In the diagnosis setting, the system generates pixel-level lesion masks and structured radiology reports. Radiologists can refine both outputs: masks can be manually adjusted at the pixel level, and report items can be modified via selectable options.

### 5.2 Workflow Integration

PASTA-AID supports two primary clinical use cases: (1) screening-aid, enabling rapid tumour detection under time-limited non-contrast CT conditions; and (2) diagnosis-aid, performing detailed lesion segmentation and automatic structured reporting on contrast-enhanced CT. In both workflows, the system emphasizes collaboration with radiologists, ensuring that users remain the final decision-makers.

### 5.3 Validation Settings

The system was evaluated in a retrospective clinical trial (ChiCTR2500101081). Radiologists of varying experience levels interacted with PASTA-AID on a 24-inch 1080p display monitor, consistent with clinical reading environments. Performance was assessed under both screening-aid and diagnosis-aid settings, with metrics covering efficiency, accuracy, and inter-observer concordance. Detailed results are reported in the Results section.

## 6. Non-contrast CT Tumour Identification

Detecting tumours in non-contrast CT scans demonstrates the ability of models to identify subtle lesions, showcasing their potential for clinical applications. This study includes 3 tumour classification tasks and 3 tumour segmentation tasks, focusing on liver cancer, pancreatic cancer, and kidney cancer. A total of 323 liver tumour scans with 916 healthy controls, 338 pancreatic tumour scans with 974 healthy controls, and 340 kidney tumour scans with 973 healthy controls were collected from the First Hospital of CMU. All CT scans are reoriented such that the first axis runs from right to left, the second from anterior to posterior, and the third from inferior to superior. The voxel spacing is then uniformly resampled to 1 mm in all directions using linear spline interpolation. For the tumour classification tasks, a conventional two-stage detection paradigm was adopted. This involved pre-extraction of target organs followed by classification of the extracted ROI volumes. TotalSegmentator was applied to segment the target organs in each scan, and the segmented regions were cropped with a 24-voxel margin in all directions. The cropped volumes were resized and padded to a uniform size of 128×128×128 voxels. The PASTA model retained its encoder and MLP components, with the MLP's final layer adjusted to output two channels (non-cancer and cancer). Models Genesis and SuPreM retained their encoders, with a three-layer MLP appended, comprising two hidden layers (channel sizes of 512 and 256) and a final output layer with two channels; each intermediate layer was followed by a LeakyReLU activation function. The training process utilized the Adam optimizer on 4 NVIDIA GTX 4090 GPU with a total batch size of 32. The training was conducted with a base learning rate of $1\times10^{-4}$ over 10000 iterations, with each epoch consisting of 250 iterations. Performance evaluation was conducted using the AUC metric with 5-fold cross-validation.

## 7. Clinical efficiency of PASTA-AID in Non-contrast CT Tumour Identification

In the PASTA-AID assisted non-contrast CT tumour identification experiments, an additional 60 tumour cases and 300 healthy organ scans were collected for each cancer type (liver, pancreas, and kidney) from the First Hospital of CMU. For each organ, the target region was first segmented using TotalSegmentator and then cropped with a 24-voxel margin in all directions to obtain the corresponding organ patch. The dataset was evenly divided between the PASTA-AID – assisted and unaided (solo) settings, ensuring no case overlap across conditions. Accordingly, for each cancer type, every radiologist was randomly assigned 30 tumour cases and 150 healthy cases in the solo setting, while the remaining cases of equal number were allocated to the PASTA-AID assisted setting. In practice, each radiologist was instructed to review as many cases as possible within a 30-minute session for each condition, rather than being required to complete the entire set. Two junior and two senior radiologists participated in the study, simulating real-world high-workload screening scenarios.

For the assisted setting, the corresponding PASTA-trained classification models were applied with 5-fold ensemble predictions, and the estimated malignancy probabilities were displayed to the radiologists. Performance was evaluated to quantify the benefit of PASTA-AID in accelerating tumour detection under high-throughput conditions, using three key metrics: 1) Recall: the proportion of true positives identified within 30 minutes relative to the total number of positive cases; 2) Precision: the proportion of true positives among all positive predictions within 30 minutes; 3) Workload throughput: the total number of cases reviewed by each radiologist within the time limit.

## 8. Full-Data Lesion Segmentation

From the lesion reference dataset, we constructed 15 lesion segmentation tasks (eTable 5), including MSD dataset[19] (lung tumour, liver tumour, pancreatic tumour and cyst, colon cancer) and the KiTS23[20] dataset (kidney tumour and cyst), as well as private scans from the First Hospital of CMU, covering liver cysts, gallbladder cancer, gallstones, esophageal cancer, gastric cancer, kidney stones, bladder cancer, and bone metastases. For the MSD Pancreas dataset, the original annotations for pancreatic cancer and pancreatic cysts shared the same label value. To address this, a senior radiologist manually differentiated and re-annotated them. For PASTA, Models Genesis, and SuPreM, we fine-tuned their encoder-decoder part using the nnUNet training framework with a base learning rate of $1\times10^{-3}$ for 500 epochs, with each epoch comprising 250 iterations. Following the default nnUNet settings, the baseline nnUNet model was trained with a learning rate of $1\times10^{-3}$ for 1,000 epochs. For the Universal model, we adhered to its standard configuration, training it with a learning rate of $1\times10^{-4}$ for 2,000 epochs. Training random-crop patch-size is 128×128×128 voxel. All experiments were conducted on a single NVIDIA GTX 4090 GPU with a batch size of 2. The performances were evaluated in terms of the DSC using the 5-fold cross-validation.

## 9. Few-shot Lesion Segmentation

Few-shot learning is a common approach for evaluating the quality of features extracted by a pre-trained model and its ability to quickly adapt to new tasks with limited annotated data and transferring iterations. In this study, the few-shot lesion segmentation experiment was designed using the same 15 segmentation tasks and datasets as described in the fully supervised lesion segmentation section (eTable 5). To simulate few-shot settings, the number of labeled training scans per task was limited to $K \in \{1,2,4,8,16\}$, and the total training iterations for all models were capped at 2,000. For PASTA, Models Genesis, and SuPreM, we fine-tuned their UNet part using the nnUNet training framework with a base learning rate of $1 \times 10^{-4}$. The baseline nnUNet model was trained with a learning rate of $1 \times 10^{-3}$. For the Universal model, the learning rate is set to $1 \times 10^{-4}$. Training random-crop patch-size is 128×128×128 voxel. All experiments were conducted on a single NVIDIA GTX 4090 GPU with a batch size of 2. The same 5-fold cross-validation splits used in the fully supervised setting were adopted, with $K$ training samples randomly selected for each fold, repeated over five runs to ensure robustness. For the KiTS23 dataset, training samples were specifically chosen to ensure that each selected scan included both kidney tumours and kidney cysts.

**10. Clinical Efficiency of PASTA-AID in Lesion Segmentation**

For the segmentation-aid scenario, 200 patients were collected for each of the three cancer types (liver, pancreas, and kidney) from the First Hospital of CMU. Two junior and two senior radiologists participated in the study. In the PASTA-AID setting, a contrast-enhanced CT scan was randomly selected for each patient, and lesion masks were first generated by an ensemble of five PASTA-based models trained in the full-data lesion segmentation tasks using 5-fold cross-validation. Radiologists then refined these auto-generated masks to obtain the final annotations.

Segmentation time was defined as the period from loading the CT scan to completion of annotation, and was automatically recorded by a timer embedded in the system. The average time for junior and senior radiologists under both solo and PASTA-AID–assisted conditions was compared to evaluate the efficiency gain provided by the system.

To evaluate the empowering effect of PASTA-AID on less-experienced radiologists, segmentations produced by junior radiologists were compared with the reference standard, defined as the final annotations reviewed and confirmed by senior radiologists. Agreement between junior and senior annotations was quantified using the Dice similarity coefficient (DSC), which measures the spatial overlap between corresponding lesion masks.

## 11. Structured Lesion Report Generation

Evaluation of structured lesion reports utilized a reference dataset covering five attributes: shape, density, heterogeneity, surface, and invasion. The prediction tasks were framed as multi-label classification problems, requiring the model to predict an option for each attribute. CT scans are standardized by orienting the first axis from right to left, the second from anterior to posterior, and the third from inferior to superior. The voxel spacing is subsequently resampled to 1 mm in all directions using linear spline interpolation. For all tasks, regions of size 96×96×96 voxels, centered on the lesion, are extracted and input into the network for classification. Fine-tuning was performed on PASTA, Models Genesis, and SuPreM models, each using a base learning rate of $1\times10^{-4}$ For Models Genesis and SuPreM, encoders were retained, with a three-layer MLP appended for attribute prediction. The training process involved Cross-Entropy Loss and the Adam optimizer with a total batch size of 32. Performance was assessed using the accuracy and F1-score metrics and 5-fold cross-validation to ensure robust evaluation.

## 12. Clinical efficiency of PASTA-AID in Structured Lesion Report Generation

For the report-generation scenario, the same cohort of 600 patients (200 each with liver, pancreatic and kidney cancers) from the First Hospital of CMU used in the segmentation-aid experiment (Method 10) was analyzed. Two junior and two senior radiologists participated, and the allocation of patients between solo and PASTA-AID–assisted conditions followed the same split scheme as in the segmentation study.

In the solo condition, radiologists reviewed contrast-enhanced CT scans and completed structured lesion reports manually. In the assisted condition, the scans were first processed by the PASTA-based encoder–classifier model described in the structured lesion report generation task (eMethod 11), which automatically produced draft reports for five attributes (shape, density, heterogeneity, surface and invasion). Radiologists then reviewed and edited these drafts to finalize the reports.

Report-generation time was defined as the interval from loading a case in the system to completion of the finalized structured report and was automatically recorded by a built-in timer. Efficiency gains were quantified by comparing average report-generation times between solo and assisted conditions for junior and senior radiologists.

To assess the impact of PASTA-AID on reporting quality, assisted reports were compared with a reference standard, defined as the final attribute annotations reviewed and confirmed by senior radiologists. Performance was evaluated using per-attribute accuracy and F1-score.

**13. Tumour Staging and Survival Predictions**

This study addresses seven tasks, including four cancer survival prediction tasks and three cancer staging tasks. The Lung1 dataset[21] consists of 422 pretreatment CT scans from non-small cell lung cancer patients, annotated with primary gross tumour volumes and overall survival information. Patients were classified into two categories: OS less than two years and OS of two years or more. The CMU gastric cancer dataset comprises 412 contrast-enhanced pretreatment abdominal CT scans from gastric cancer patients, annotated with tumour center points, survival data, and overall staging. The survival prediction task divides patients into OS less than two years and OS of two years or more, while staging distinguishes between stages I-II and III-IV. Similarly, the CMU rectal cancer dataset includes 160 contrast-enhanced pretreatment abdominal CT scans annotated with tumour center points, survival information, and overall staging, with tasks involving OS less than three years and OS of three years or more, as well as staging differentiation between stages I-III and stage IV. The TCGA-BLCA[22] dataset features 120 bladder cancer patients with imaging data and survival information, complemented by radiologist-annotated tumour center points. Survival prediction categorizes patients into OS less than three years and OS of three years or more, while staging separates stages I-II from III-IV. All CT scans are reformatted with the first axis oriented from right to left, the second from anterior to posterior, and the third from inferior to superior. The spacing in all directions is then resampled to 1 mm using linear spline interpolation. For all tasks, the FMCIB pipeline was adopted, which involves feature extraction from ROI regions centered on annotated tumour points for classification. Fine-tuning was performed on PASTA, Models Genesis, SuPreM, and FMCIB models, each using a base learning rate of $1\times10^{-4}$. For Models Genesis, SuPreM, and FMCIB models, encoders were retained, with a three-layer MLP appended for attribute prediction. The training process involved Cross-Entropy Loss and the Adam optimizer with a total batch size of 32. Performance was assessed using the AUC metric and 5-fold cross-validation to ensure robust evaluation.

**14. Efficient Oncology Transfer Learning Across Modalities**

A cross-modality evaluation was performed on two cancer segmentation tasks: (1) the MSD-Brain Tumours segmentation task[25], comprising 484 patients with FLAIR, T1w, T1gd, and T2w MRI scans. The segmentation targets included necrotic/active tumour regions and edema in gliomas. (2) The ATLAS dataset[23], which includes 60 contrast-enhanced T1-weighted MRI scans of the liver from 60 patients with unresectable hepatocellular carcinoma (HCC), along with segmentation masks for the liver and liver tumours. In a few-shot learning setup, the number of labeled training scans was limited to $K\in\{1,2,4,8,16\}$, and total training iterations were capped at 2,000. PASTA, Models Genesis, and SuPreM models were fine-tuned using their UNet components within the nnUNet training framework, employing a base learning rate of $1\times10^{-4}$. The nnUNet baseline model was trained with a learning rate of $1\times10^{-3}$, while the Universal model used a learning rate of $1\times10^{-4}$. The training employed random-cropped patches of 128×128×128 voxels. All experiments were performed on a single NVIDIA GTX 4090 GPU with a batch size of 2. The 5-fold cross-validation splits from the fully supervised setting were applied. The 5-fold cross-validation splits from the fully supervised setting were applied. For each fold, $K$ training samples were randomly selected from the training set, and this process was repeated across five runs.

**15. Data Availability**

To advance oncology research, we release the PASTA-Gen-30K dataset, which consists of 30,000 synthetic 3D-CT scans with corresponding lesion masks and structured report descriptions at https://huggingface.co/datasets/LWHYC/PASTA-Gen-30K. The publicly available datasets used in this study can be accessed through the following sources: MSD (http://medicaldecathlon.com/); KiTS23 (https://kits-challenge.org/kits23/); Lung1 (https://www.cancerimagingarchive.net/collection/nsclc-radiomics/); TCGA-BLCA (https://www.cancerimagingarchive.net/collection/tcga-blca/); ATLAS (https://atlas-challenge.u-bourgogne.fr/). Structured report annotations for publicly available datasets, as well as lesion center point annotations for Lung1 and TCGA-BLCA, can be accessed at https://github.com/LWHYC/PASTA. Additional imaging data and associated clinical records were obtained from the First Hospital of China Medical University. Due to privacy regulations, these data cannot be publicly shared. This retrospective clinical study was reviewed and registered in the Chinese Clinical Trial Registry (https://www.chictr.org.cn/) under the registration number ChiCTR2500101081.

**16. Code Availability**

The PASTA model, along with the full training and evaluation pipeline, is publicly available to facilitate further research and replication of our findings. The source code, model weights, and implementation details can be accessed at https://github.com/LWHYC/PASTA, ensuring accessibility for the broader clinical and scientific community.

**eFigures**

**eFigure 1: Overview of denoising network training process.**

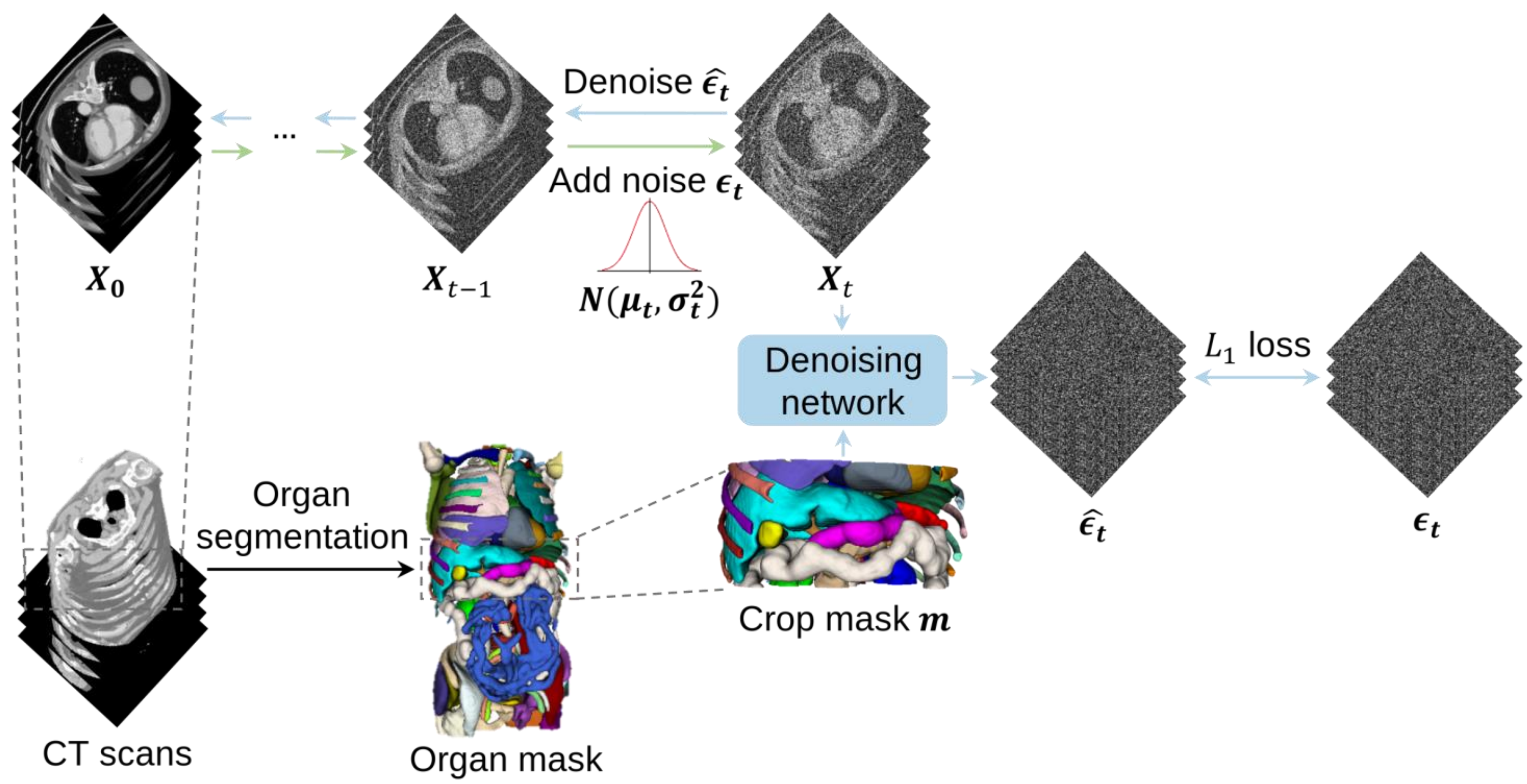

**Overview of denoising network training process.** During training, random Gaussian noise is added to the cropped original CT images ($X_0$), and the network learns to predict and remove this noise. The corresponding organ segmentation mask is concatenated as an input condition to guide the denoising process.

**eFigure 2: Evaluation of Image Realism and Description Accuracy for PASTA-Gen.**

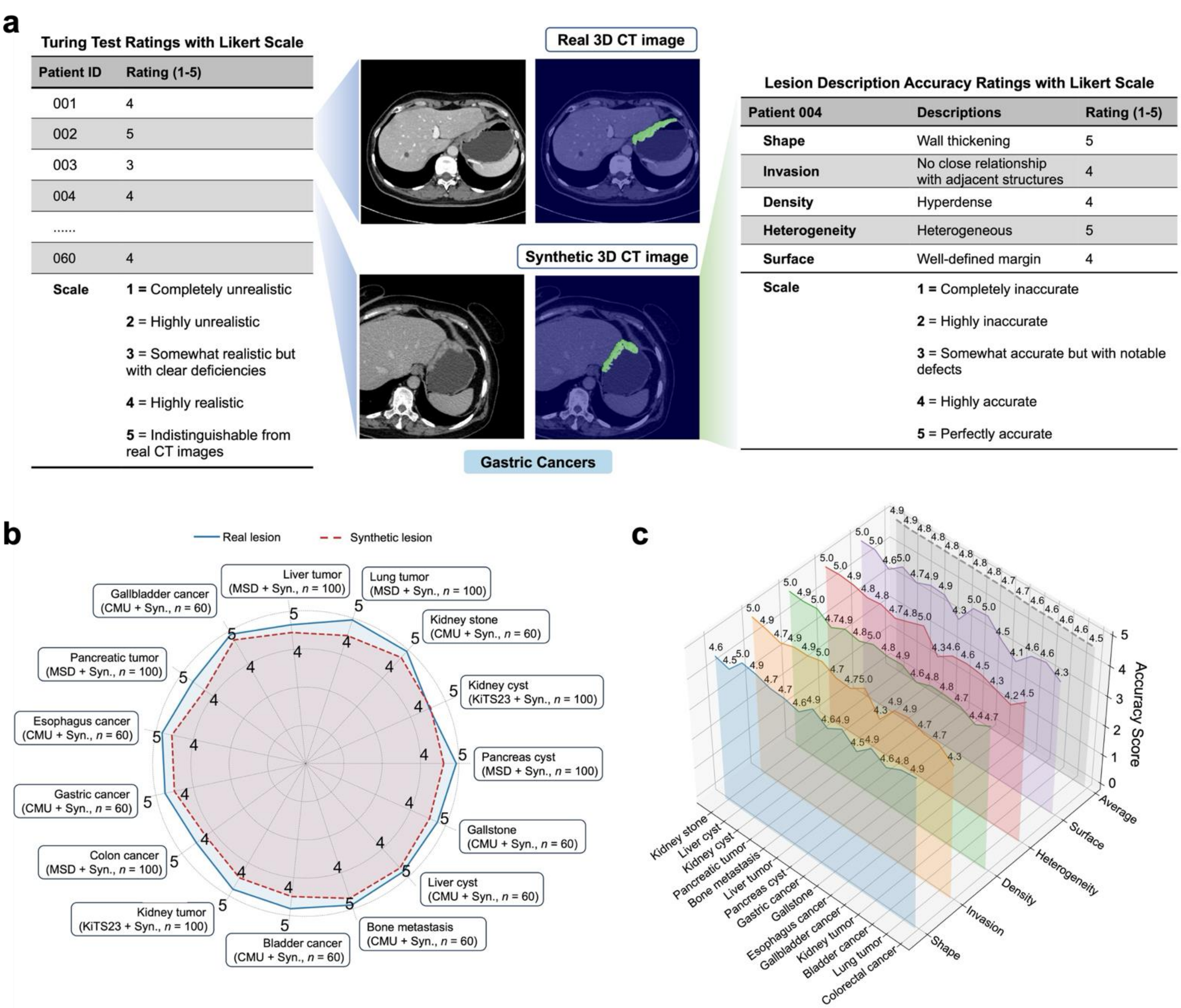

**Evaluation of Image Realism and Description Accuracy for PASTA-Gen. a**, Illustration of the Turing test for image realism and the assessment of lesion description accuracy, both rated on a Likert scale. **b**, Average Turing test scores of real and synthetic lesions. **c**, Average accuracy scores of structured report descriptions for each lesion type.

**eFigure 3 (Part 1): Examples from PASTA-Gen-30K (Part 1)**

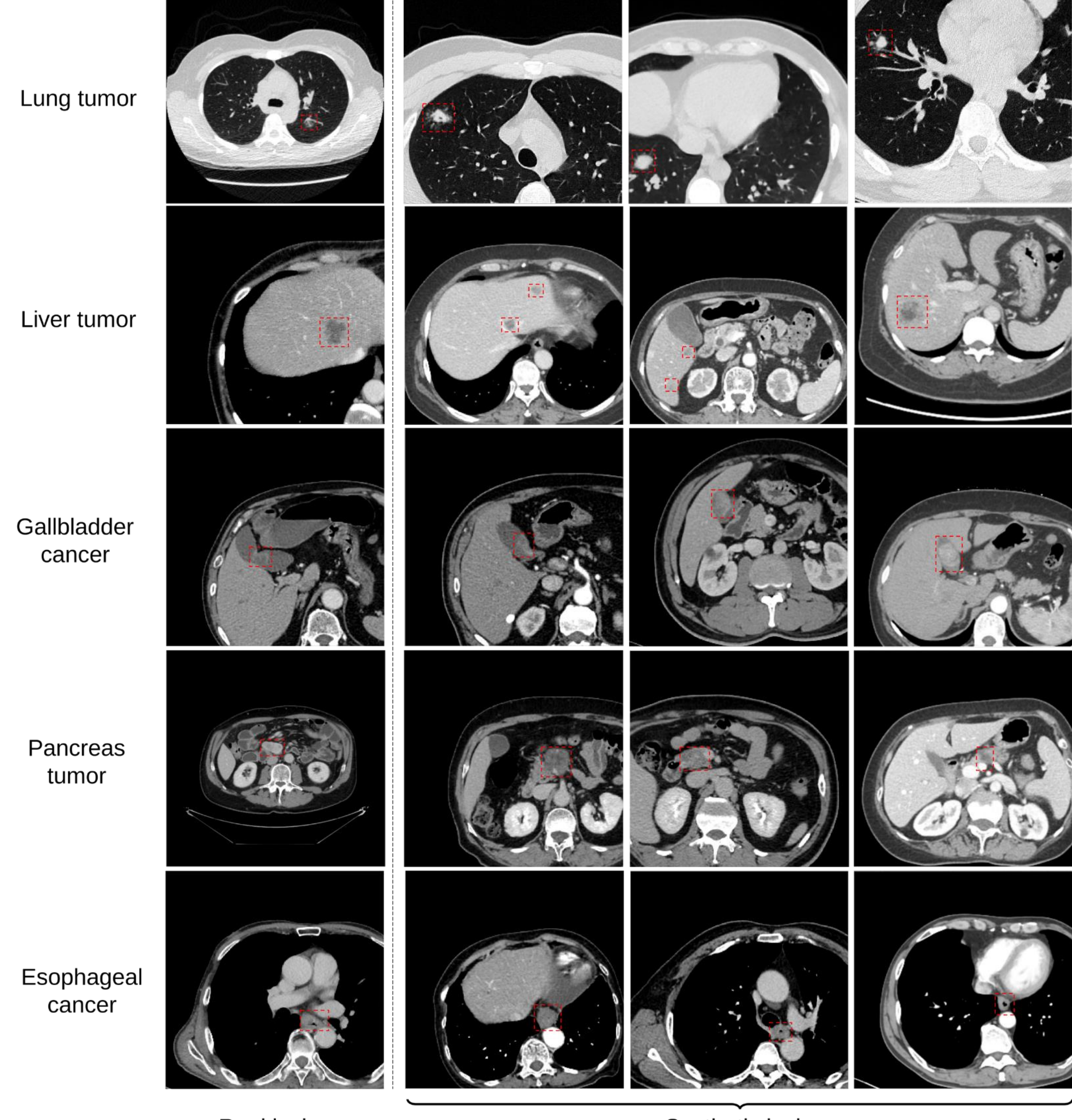

**eFigure 3 (Part 2): Examples from PASTA-Gen-30K (Part 2)**

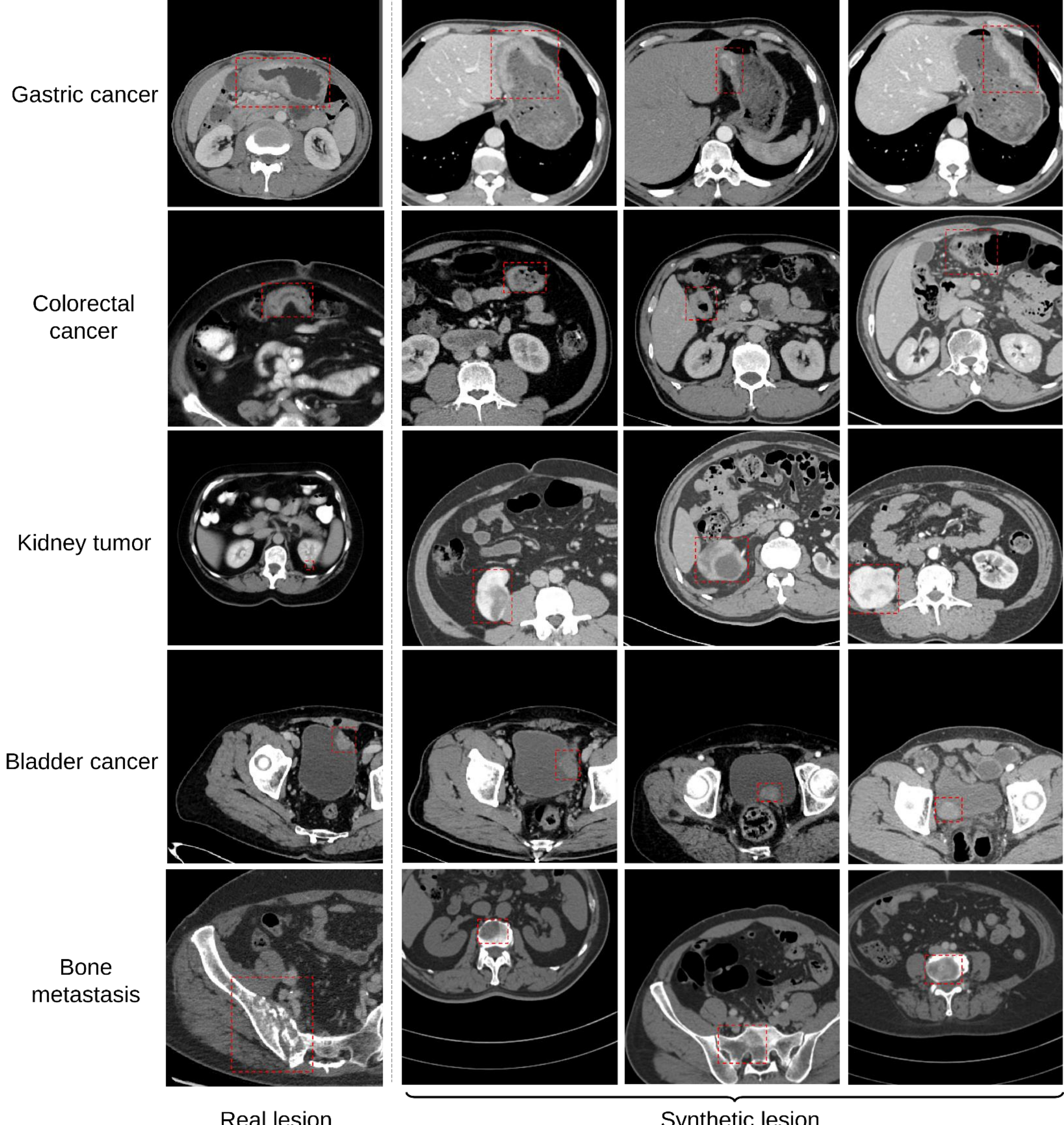

**eFigure 3 (Part 3): Examples from PASTA-Gen-30K (Part 3)**

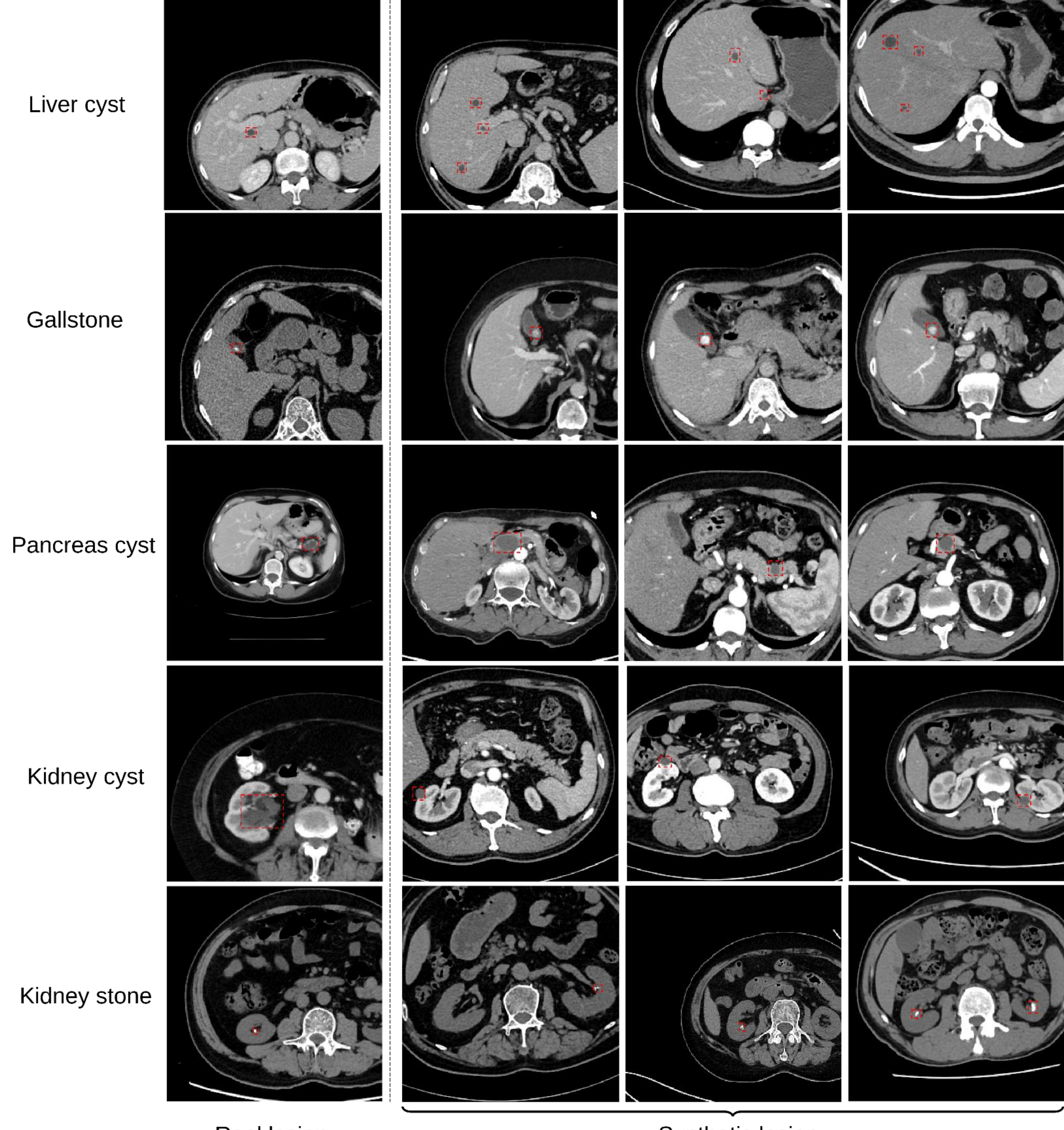

**eTables**

**eTable 1: Structured Attributes for Lesion Simulation in PASTA-Gen**

| Attribute | Definition | Structured descriptions/options |
|---|---|---|
| Enhancement status | Intravenous contrast agent usage | "Enhanced CT", "Non-contrast CT" |
| Location | Organ-specific anatomical regions | "Lung", "Liver", "Gallbladder", "Pancreas", "Esophagus", "Stomach", "Colorectal", "Kidney", "Bladder", "Bone" |
| Size | Dimensions across axial, sagittal, and coronal planes | Z×X×Y mm |
| Shape | Morphological characteristics of the lesion | "Round-like", "Irregular", "Wall thickening", "Punctate, nodular" |
| Density | Radiographic attenuation properties of the lesion | "Hypodense", "Isodense", "Hyperdense" |
| Heterogeneity | Uniformity of attenuation within the lesion | "Homogeneous", "Heterogeneous" |
| Surface | Features of the lesion's border and surface texture | "Well-defined margin", "Ill-defined margin" |
| Invasion | Invasion of or proximity to adjacent structures | "No close relationship with surrounding structures", "Close relationship with adjacent structures" |

**eTable 2: Distribution of Scan Ranges in the CMU In-house Dataset.**

| **Scan Range** | **No. of scans (%)** |
|---|---|
| Thorax CT | 5,642 (47.5%) |
| Abdomen-pelvis CT | (16.4%) |
| Thorax-abdomen-pelvis CT | 763 (7.8%) |
| Others | 2,761 (28.3%) |
| **Total** | 10,767 (100.0%) |

**eTable 3: Number of Healthy Organ Scans in the CMU Template CT Dataset (CMU-TE)**

| Organ | No. of scans |
| --- | --- |
| **Healthy Lung** | 1,720 |
| Contrast CT | 0 |
| Non-contrast CT | 1,720 |
| **Healthy Liver** | 816 |
| Contrast CT | 600 |
| Non-contrast CT | 216 |
| **Healthy Gallbladder** | 1,901 |
| Contrast CT | 1,389 |
| Non-contrast CT | 512 |
| **Healthy Pancrease** | 2,222 |
| Contrast CT | 1,620 |
| Non-contrast CT | 602 |
| **Healthy Esophagus** | 752 |
| Contrast CT | 545 |
| Non-contrast CT | 207 |
| **Healthy Stomach** | 2,281 |
| Contrast CT | 1,664 |
| Non-contrast CT | 617 |
| **Healthy Colorectum** | 1,668 |
| Contrast CT | 1,292 |
| Non-contrast CT | 376 |
| **Healthy Kidney** | 975 |
| Contrast CT | 716 |
| Non-contrast CT | 259 |
| **Healthy Bladder** | 2,216 |
| Contrast CT | 1,616 |
| Non-contrast CT | 600 |
| **Healthy Trunk and Extremity Bones** | 6,811 |
| Contrast CT | 4,972 |
| Non-contrast CT | 1,839 |
| **Total** | 21,362 |

**eTable 4: Composition of the CMU Lesion Simulation Reference Dataset (CMU-LSR)**

| Lesion targets | No. of CT scans |
|---|---|
| Lung tumour | 50 |
| Liver tumour | 50 |
| Gallbladder cancer | 50 |
| Pancreatic tumour | 50 |
| Esophageal Cancer | 50 |
| Gastric cancer | 50 |
| Colorectal cancer | 50 |
| Kidney tumour | 50 |
| Bladder cancer | 50 |
| Bone metastasis | 50 |
| Liver cyst | 50 |
| Gallstone | 50 |
| Pancreatic cyst | 50 |
| Kidney cyst | 50 |
| Kidney stone | 50 |
| **Total** | 750 |

**eTable 5: Segmentation and Structured Report Generation Datasets**

| Dataset name | Lesion target | No. of scans |
|---|---|---|
| **Public datasets** | | |
| KiTS23 | Kidney tumour, kidney cyst | 489 |
| MSD-Colon Tumour | Colon tumour | 126 |
| MSD-Liver Tumour | Liver tumour | 303 |
| MSD-Lung Tumour | Lung tumour | 96 |
| MSD-Pancreas* | Pancreatic tumour | 216 |
| MSD-Pancreas* | Pancreatic cyst | 65 |
| MSD-Brain Tumour | Necrotic tumour regions, active tumour regions, edema in gliomas. | 484 |
| ATLAS | Liver tumour | 60 |
| **CMU-LSSR datasets** | Liver cyst | 30 |
| | Gallbladder cancer | 30 |
| | Gallstones | 30 |
| | Esophageal cancer | 30 |
| | Gastric cancer | 30 |
| | Kidney stone | 30 |
| | Bladder cancer | 30 |
| | Bone metastasis | 30 |
| **Total** | | 2,089 |

* MSD-Pancreas is reannotated to pancreatic tumour and pancreatic cyst.

**eTable 6: Composition of the Image Realism Evaluation Dataset**

| Lesion targets | No. of real CT scans | Real CT sources | No. of synthetic CT scans |
|---|---|---|---|
| Lung tumour | 50 | MSD | 50 |
| Liver tumour | 50 | MSD | 50 |
| Gallbladder cancer | 30 | CMU-LSSR | 30 |
| Pancreatic tumour | 50 | MSD | 50 |
| Esophageal Cancer | 30 | CMU-LSSR | 30 |
| Gastric cancer | 30 | CMU-LSSR | 30 |
| Colorectal cancer | 50 | MSD | 50 |
| Kidney tumour | 50 | KiTS23 | 50 |
| Bladder cancer | 30 | CMU-LSSR | 30 |
| Bone metastasis | 30 | CMU-LSSR | 30 |
| Liver cyst | 30 | CMU-LSSR | 30 |
| Gallstone | 30 | CMU-LSSR | 30 |
| Pancreatic cyst | 50 | MSD | 50 |
| Kidney cyst | 50 | KiTS23 | 50 |
| Kidney stone | 30 | CMU-LSSR | 30 |
| **Total** | 590 | | 590 |

**eTable 7: Comparison of Tumour Detection on Non-contrast CT Data Based on AUC.**

| | PASTA | PASTA (w/o pretraining) | Models Genesis | SuPerM | *p*-value* |
|---|---|---|---|---|---|
| **Liver cancer** | **0.965 (0.950-0.985)** | 0.759 (0.737-0.804) | 0.765 (0.725-0.823) | 0.855 (0.817-0.911) | 0.031 |
| **Pancreatic cancer** | **0.987 (0.981-0.995)** | 0.597 (0.588-0.668) | 0.774 (0.725-0.860) | 0.944 (0.822-0.983) | 0.031 |
| **Kidney cancer** | **0.964 (0.954-0.983)** | 0.781 (0.743-0.836) | 0.705 (0.767-0.800) | 0.908 (0.870-0.942) | 0.031 |

The values in parentheses represent the minimum and maximum AUC values obtained across the 5-fold cross-validation.

* *p*-values were calculated using a one-sided Wilcoxon test comparing the **top-performing model** and the second-best model for each task

**eTable 8: Performance of Radiologists in Non-contrast CT Tumour Screening with and without PASTA-AID Assistance**

| | Liver Cancer | | | | Pancreatic cancer | | | | Kidney Cancer | | | |
|---|---|---|---|---|---|---|---|---|---|---|---|---|
| | TP | TN | FP | FN | TP | TN | FP | FN | TP | TN | FP | FN |
| Junior Radiologist 1 | | | | | | | | | | | | |
| Solo | 5 | 37 | 3 | 3 | 2 | 24 | 1 | 4 | 5 | 45 | 2 | 4 |
| PASTA-AID | 9 | 63 | 0 | 1 | 10 | 31 | 1 | 0 | 11 | 61 | 1 | 0 |
| Junior Radiologist 2 | | | | | | | | | | | | |
| Solo | 7 | 38 | 2 | 3 | 5 | 27 | 0 | 5 | 6 | 44 | 1 | 4 |
| PASTA-AID | 10 | 59 | 1 | 1 | 9 | 30 | 0 | 2 | 10 | 68 | 0 | 2 |
| Senior Radiologist 1 | | | | | | | | | | | | |
| Solo | 11 | 50 | 4 | 2 | 9 | 42 | 2 | 2 | 14 | 65 | 2 | 3 |
| PASTA-AID | 15 | 56 | 1 | 1 | 8 | 44 | 1 | 1 | 14 | 72 | 1 | 1 |
| Senior Radiologist 2 | | | | | | | | | | | | |
| Solo | 10 | 53 | 4 | 3 | 8 | 53 | 3 | 2 | 12 | 59 | 3 | 4 |
| PASTA-AID | 14 | 61 | 0 | 2 | 10 | 61 | 2 | 0 | 13 | 68 | 0 | 2 |
| Avg. Cases Read (Solo) | 58.8 | | | | 47.3 | | | | 68.3 | | | |
| Avg. Cases Read (PASTA-AID) | 73.5 | | | | 52.5 | | | | 81.0 | | | |
| Efficiency Improvement (%) | 25.1 | | | | 11.1 | | | | 18.7 | | | |
| Avg. Recall (Solo, %) | 73.5 | | | | 61.3 | | | | 68.2 | | | |
| Avg. Recall (PASTA-AID, %) | 90.5 | | | | 92.7 | | | | 90.8 | | | |
| Recall Improvement (%) | 17.0 | | | | 31.4 | | | | 22.6 | | | |
| *p-value for recall* | **0.043** | | | | **0.002** | | | | **0.010** | | | |
| Avg. Precsion (Solo, %) | 71.3 | | | | 80.3 | | | | 81.2 | | | |
| Avg. Precsion (PASTA-AID, %) | 96.2 | | | | 90.8 | | | | 96.3 | | | |
| Precision Improvement (%) | 24.9 | | | | 10.5 | | | | 15.1 | | | |
| *p-value for precision* | **0.001** | | | | 0.279 | | | | **0.027** | | | |

TP, true positive; TN, true negative; FP, false positive; FN, false negative.

Reading time for each round was limited to 30 minutes.

*p*-values were calculated using a two-sided Cochran–Mantel–Haenszel test stratified by reader across four radiologists, comparing PASTA-AID–assisted versus solo readings for each diagnostic metric.

**eTable 9: Sufficient Data and Few-Shot Segmentation DSC across Models.**

| | PASTA | Models Genesis | SuPerM | nnUNet | Universal | *p*-value* |
|---|---|---|---|---|---|---|
| **Lung tumour** | | | | | | |
| No. of Few-shot Samples† | | | | | | |
| 1 | **0.529 (0.458-0.599)** | 0.179 (0.127-0.230) | 0.302 (0.230-0.383) | 0.188 (0.133-0.254) | 0.167 (0.118-0.219) | <0.001 |
| 2 | **0.622 (0.559-0.683)** | 0.358 (0.291-0.425) | 0.383 (0.310-0.457) | 0.288 (0.220-0.360) | 0.266 (0.203-0.330) | <0.001 |
| 4 | **0.680 (0.621-0.728)** | 0.396 (0.321-0.464) | 0.513 (0.444-0.590) | 0.335 (0.265-0.405) | 0.307 (0.247-0.378) | <0.001 |
| 8 | **0.709 (0.656-0.757)** | 0.365 (0.309-0.423) | 0.605 (0.546-0.660) | 0.323 (0.257-0.389) | 0.293 (0.233-0.361) | <0.001 |
| 16 | **0.706 (0.651-0.751)** | 0.261 (0.206-0.322) | 0.591 (0.531-0.654) | 0.106 (0.073-0.142) | 0.088 (0.059-0.121) | <0.001 |
| Full-scale data training (n=64) | **0.708 (0.639-0.767)** | 0.680 (0.616-0.735) | 0.661 (0.595-0.720) | 0.689 (0.630-0.744) | 0.661 (0.602-0.708) | 0.018 |
| **Liver tumour** | | | | | | |
| No. of Few-shot Samples† | | | | | | |
| 1 | **0.243 (0.199-0.290)** | 0.101 (0.078-0.129) | 0.194 (0.148-0.247) | 0.097 (0.070-0.125) | 0.080 (0.057-0.106) | 0.004 |
| 2 | **0.477 (0.430-0.527)** | 0.145 (0.112-0.176) | 0.313 (0.258-0.369) | 0.164 (0.133-0.202) | 0.141 (0.110-0.176) | <0.001 |
| 4 | **0.598 (0.548-0.643)** | 0.193 (0.157-0.233) | 0.392 (0.341-0.451) | 0.170 (0.134-0.206) | 0.152 (0.121-0.186) | <0.001 |
| 8 | **0.617 (0.568-0.658)** | 0.108 (0.081-0.138) | 0.390 (0.334-0.447) | 0.064 (0.041-0.090) | 0.060 (0.040-0.084) | <0.001 |
| 16 | **0.629 (0.580-0.677)** | 0.131 (0.100-0.164) | 0.316 (0.261-0.376) | 0.018 (0.011-0.025) | 0.012 (0.008-0.016) | <0.001 |
| Full-scale data training (n=131) | **0.696 (0.652-0.738)** | 0.673 (0.627-0.716) | 0.654 (0.607-0.701) | 0.666 (0.619-0.713) | 0.640 (0.591-0.687) | <0.001 |
| **Gallbladder cancer** | | | | | | |
| No. of Few-shot Samples† | | | | | | |
| 1 | **0.565 (0.454-0.673)** | 0.112 (0.062-0.171) | 0.065 (0.009-0.131) | 0.101 (0.051-0.160) | 0.096 (0.042-0.152) | <0.001 |
| 2 | **0.608 (0.509-0.696)** | 0.122 (0.067-0.179) | 0.145 (0.074-0.226) | 0.104 (0.055-0.162) | 0.095 (0.046-0.153) | <0.001 |
| 4 | **0.649** | 0.079 | 0.217 | 0.152 | 0.130 | <0.001 |
| | **(0.554-0.736)** | (0.039-0.125) | (0.137-0.301) | (0.097-0.216) | (0.080-0.190) | |
| 8 | **0.630**<br>**(0.528-0.721)** | 0.206<br>(0.128-0.286) | 0.419<br>(0.317-0.518) | 0.203<br>(0.137-0.269) | 0.181<br>(0.116-0.242) | <0.001 |
| 16 | **0.653**<br>**(0.544-0.748)** | 0.107<br>(0.059-0.158) | 0.365<br>(0.268-0.465) | 0.105<br>(0.065-0.148) | 0.081<br>(0.047-0.116) | <0.001 |
| Full-scale data training (n=30) | **0.654**<br>**(0.556-0.744)** | 0.598<br>(0.484-0.700) | 0.575<br>(0.470-0.671) | 0.644<br>(0.541-0.742) | 0.613<br>(0.517-0.708) | 0.464 |
| **Pancreatic tumour** | | | | | | |
| No. of Few-shot Samples† | | | | | | |
| 1 | **0.158**<br>**(0.133-0.186)** | 0.037<br>(0.024-0.052) | 0.041<br>(0.027-0.058) | 0.046<br>(0.032-0.062) | 0.040<br>(0.026-0.054) | <0.001 |
| 2 | **0.216**<br>**(0.187-0.252)** | 0.073<br>(0.054-0.092) | 0.079<br>(0.060-0.101) | 0.055<br>(0.038-0.075) | 0.051<br>(0.035-0.065) | <0.001 |
| 4 | **0.209**<br>**(0.175-0.244)** | 0.015<br>(0.007-0.026) | 0.122<br>(0.096-0.150) | 0.005<br>(0.002-0.010) | 0.007<br>(0.004-0.010) | <0.001 |
| 8 | **0.335**<br>**(0.294-0.375)** | 0.043<br>(0.029-0.057) | 0.141<br>(0.113-0.169) | 0.000<br>(0.000-0.000) | 0.005<br>(0.004-0.006) | <0.001 |
| 16 | **0.414**<br>**(0.374-0.454)** | 0.066<br>(0.049-0.085) | 0.207<br>(0.176-0.241) | 0.000<br>(0.000-0.000) | 0.005<br>(0.004-0.007) | <0.001 |
| Full-scale data training (n=220) | **0.510**<br>**(0.468-0.551)** | 0.482<br>(0.442-0.520) | 0.408<br>(0.365-0.449) | 0.491<br>(0.452-0.531) | 0.468<br>(0.428-0.508) | 0.010 |
| **Esophageal Cancer** | | | | | | |
| No. of Few-shot Samples† | | | | | | |
| 1 | **0.577**<br>**(0.466-0.683)** | 0.166<br>(0.089-0.253) | 0.319<br>(0.218-0.419) | 0.175<br>(0.089-0.268) | 0.163<br>(0.089-0.248) | <0.001 |
| 2 | **0.608**<br>**(0.507-0.698)** | 0.279<br>(0.184-0.377) | 0.422<br>(0.324-0.518) | 0.351<br>(0.254-0.458) | 0.323<br>(0.232-0.417) | <0.001 |
| 4 | **0.620**<br>**(0.520-0.719)** | 0.458<br>(0.357-0.552) | 0.384<br>(0.255-0.512) | 0.350<br>(0.242-0.464) | 0.312<br>(0.214-0.415) | <0.001 |
| 8 | **0.659**<br>**(0.555-0.758)** | 0.345<br>(0.234-0.460) | 0.491<br>(0.376-0.604) | 0.506<br>(0.405-0.605) | 0.476<br>(0.389-0.566) | <0.001 |
| 16 | **0.684**<br>**(0.582-0.775)** | 0.385<br>(0.275-0.500) | 0.624<br>(0.527-0.709) | 0.562<br>(0.481-0.637) | 0.526<br>(0.447-0.604) | 0.007 |
| Full-scale data training (n=30) | **0.683**<br>**(0.576-0.783)** | 0.635<br>(0.508-0.741) | 0.646<br>(0.535-0.746) | 0.634<br>(0.514-0.739) | 0.602<br>(0.479-0.703) | 0.004 |
| **Gastric Cancer** | | | | | | |
| No. of Few-shot Samples† | | | | | | |
| 1 | **0.309**<br>**(0.209-0.418)** | 0.074<br>(0.039-0.110) | 0.045<br>(0.015-0.083) | 0.054<br>(0.021-0.091) | 0.045<br>(0.017-0.080) | <0.001 |
| 2 | **0.494** | 0.098 | 0.033 | 0.090 | 0.067 | <0.001 |
| | **(0.413-0.574)** | (0.056-0.142) | (0.013-0.056) | (0.049-0.132) | (0.031-0.112) | |
| 4 | **0.548**<br>**(0.469-0.623)** | 0.183<br>(0.129-0.241) | 0.235<br>(0.165-0.302) | 0.165<br>(0.110-0.220) | 0.135<br>(0.081-0.190) | <0.001 |
| 8 | **0.584**<br>**(0.505-0.655)** | 0.222<br>(0.159-0.293) | 0.383<br>(0.295-0.455) | 0.217<br>(0.161-0.276) | 0.185<br>(0.129-0.245) | <0.001 |
| 16 | **0.615**<br>**(0.544-0.677)** | 0.275<br>(0.219-0.330) | 0.461<br>(0.398-0.524) | 0.247<br>(0.204-0.288) | 0.203<br>(0.162-0.241) | <0.001 |
| Full-scale data training (n=30) | **0.627**<br>**(0.542-0.697)** | 0.561<br>(0.496-0.631) | 0.544<br>(0.460-0.617) | 0.581<br>(0.505-0.646) | 0.554<br>(0.486-0.624) | 0.004 |
| **Colon Cancer** | | | | | | |
| No. of Few-shot Samples† | | | | | | |
| 1 | **0.044**<br>**(0.027-0.066)** | 0.006<br>(0.000-0.015) | 0.019<br>(0.008-0.034) | 0.005<br>(0.002-0.008) | 0.008<br>(0.006-0.010) | 0.001 |
| 2 | **0.167**<br>**(0.126-0.208)** | 0.024<br>(0.012-0.041) | 0.050<br>(0.029-0.071) | 0.023<br>(0.011-0.038) | 0.026<br>(0.014-0.040) | <0.001 |
| 4 | **0.245**<br>**(0.204-0.292)** | 0.042<br>(0.024-0.064) | 0.115<br>(0.083-0.147) | 0.061<br>(0.042-0.085) | 0.050<br>(0.031-0.069) | <0.001 |
| 8 | **0.290**<br>**(0.240-0.343)** | 0.082<br>(0.059-0.107) | 0.148<br>(0.117-0.182) | 0.093<br>(0.069-0.119) | 0.077<br>(0.054-0.101) | <0.001 |
| 16 | **0.391**<br>**(0.338-0.445)** | 0.112<br>(0.086-0.142) | 0.110<br>(0.083-0.139) | 0.059<br>(0.046-0.074) | 0.043<br>(0.031-0.056) | <0.001 |
| Full-scale data training (n=126) | **0.517**<br>**(0.458-0.583)** | 0.496<br>(0.436-0.556) | 0.481<br>(0.425-0.542) | 0.517<br>(0.454-0.576) | 0.491<br>(0.438-0.547) | 0.901 |
| **Kidney tumour** | | | | | | |
| No. of Few-shot Samples† | | | | | | |
| 1 | **0.246**<br>**(0.220-0.271)** | 0.039<br>(0.030-0.049) | 0.072<br>(0.059-0.086) | 0.052<br>(0.042-0.061) | 0.043<br>(0.034-0.052) | <0.001 |
| 2 | **0.301**<br>**(0.269-0.330)** | 0.015<br>(0.010-0.019) | 0.127<br>(0.110-0.146) | 0.046<br>(0.037-0.054) | 0.037<br>(0.030-0.046) | <0.001 |
| 4 | **0.394**<br>**(0.361-0.428)** | 0.006<br>(0.003-0.010) | 0.174<br>(0.152-0.198) | 0.021<br>(0.014-0.029) | 0.021<br>(0.014-0.027) | <0.001 |
| 8 | **0.632**<br>**(0.606-0.658)** | 0.133<br>(0.116-0.150) | 0.254<br>(0.229-0.279) | 0.143<br>(0.128-0.159) | 0.124<br>(0.107-0.139) | <0.001 |
| 16 | **0.678**<br>**(0.654-0.706)** | 0.144<br>(0.129-0.161) | 0.336<br>(0.309-0.361) | 0.051<br>(0.042-0.062) | 0.046<br>(0.036-0.055) | <0.001 |
| Full-scale data training (n=489) | **0.814**<br>**(0.793-0.834)** | 0.800<br>(0.780-0.819) | 0.722<br>(0.695-0.748) | 0.784<br>(0.763-0.805) | 0.756<br>(0.736-0.776) | <0.001 |
| **Bladder cancer** | | | | | | |
| No. of Few-shot Samples† | | | | | | |
| 1 | **0.509** | 0.136 | 0.052 | 0.081 | 0.075 | <0.001 |
| | **(0.381-0.631)** | (0.064-0.213) | (0.000-0.122) | (0.018-0.158) | (0.014-0.152) | |
| 2 | **0.667 (0.566-0.764)** | 0.290 (0.196-0.396) | 0.194 (0.105-0.294) | 0.219 (0.140-0.311) | 0.201 (0.115-0.297) | <0.001 |
| 4 | **0.697 (0.600-0.781)** | 0.445 (0.355-0.535) | 0.396 (0.289-0.500) | 0.360 (0.264-0.461) | 0.332 (0.242-0.426) | <0.001 |
| 8 | **0.712 (0.611-0.797)** | 0.500 (0.400-0.588) | 0.441 (0.325-0.553) | 0.318 (0.215-0.421) | 0.289 (0.194-0.385) | <0.001 |
| 16 | **0.757 (0.674-0.828)** | 0.550 (0.466-0.637) | 0.309 (0.226-0.394) | 0.331 (0.230-0.432) | 0.301 (0.205-0.404) | <0.001 |
| Full-scale data training (n=30) | **0.790 (0.717-0.844)** | 0.769 (0.694-0.830) | 0.660 (0.550-0.756) | 0.759 (0.680-0.829) | 0.738 (0.658-0.806) | 0.127 |
| **Bone metastasis** | | | | | | |
| No. of Few-shot Samples† | | | | | | |
| 1 | **0.313 (0.266-0.386)** | 0.064 (0.015-0.120) | 0.063 (0.019-0.122) | 0.067 (0.021-0.124) | 0.065 (0.020-0.123) | <0.001 |
| 2 | **0.360 (0.251-0.421)** | 0.049 (0.001-0.118) | 0.050 (0.009-0.106) | 0.058 (0.011-0.128) | 0.050 (0.007-0.111) | <0.001 |
| 4 | **0.371 (0.287-0.458)** | 0.068 (0.023-0.123) | 0.053 (0.013-0.108) | 0.076 (0.026-0.135) | 0.064 (0.022-0.113) | <0.001 |
| 8 | **0.402 (0.298-0.485)** | 0.124 (0.059-0.205) | 0.060 (0.020-0.115) | 0.081 (0.033-0.138) | 0.074 (0.025-0.145) | <0.001 |
| 16 | **0.432 (0.267-0.538)** | 0.023 (0.010-0.040) | 0.002 (0.000-0.007) | 0.031 (0.011-0.058) | 0.023 (0.006-0.051) | <0.001 |
| Full-scale data training (n=30) | **0.433 (0.317-0.574)** | 0.321 (0.210-0.431) | 0.389 (0.284-0.489) | 0.376 (0.252-0.488) | 0.346 (0.245-0.458) | 0.020 |
| **Liver cyst** | | | | | | |
| No. of Few-shot Samples† | | | | | | |
| 1 | **0.639 (0.545-0.727)** | 0.015 (0.000-0.042) | 0.286 (0.172-0.407) | 0.000 (0.000-0.000) | 0.009 (0.004-0.015) | <0.001 |
| 2 | **0.631 (0.533-0.740)** | 0.038 (0.008-0.075) | 0.199 (0.087-0.328) | 0.000 (0.000-0.000) | 0.006 (0.003-0.010) | <0.001 |
| 4 | **0.748 (0.671-0.814)** | 0.022 (0.001-0.050) | 0.315 (0.204-0.436) | 0.000 (0.000-0.000) | 0.004 (0.001-0.008) | <0.001 |
| 8 | **0.716 (0.619-0.802)** | 0.056 (0.009-0.114) | 0.271 (0.161-0.391) | 0.000 (0.000-0.000) | 0.002 (0.000-0.006) | <0.001 |
| 16 | **0.789 (0.694-0.854)** | 0.010 (0.001-0.025) | 0.299 (0.165-0.454) | 0.000 (0.000-0.000) | 0.002 (0.000-0.005) | <0.001 |
| Full-scale data training (n=30) | **0.755 (0.649-0.846)** | 0.713 (0.611-0.799) | 0.724 (0.628-0.802) | 0.716 (0.612-0.794) | 0.686 (0.588-0.773) | <0.001 |
| **Gallstone** | | | | | | |
| No. of Few-shot | | | | | | |
| Samples† | | | | | | |
| 1 | **0.444 (0.312-0.581)** | 0.100 (0.030-0.182) | 0.072 (0.000-0.153) | 0.092 (0.022-0.185) | 0.087 (0.024-0.172) | <0.001 |
| 2 | **0.637 (0.531-0.732)** | 0.190 (0.097-0.286) | 0.254 (0.159-0.354) | 0.280 (0.164-0.396) | 0.247 (0.145-0.354) | <0.001 |
| 4 | **0.687 (0.589-0.776)** | 0.321 (0.208-0.427) | 0.364 (0.254-0.486) | 0.348 (0.244-0.454) | 0.329 (0.226-0.437) | <0.001 |
| 8 | **0.769 (0.692-0.829)** | 0.493 (0.368-0.611) | 0.366 (0.255-0.492) | 0.510 (0.397-0.619) | 0.477 (0.360-0.580) | <0.001 |
| 16 | **0.762 (0.684-0.833)** | 0.465 (0.352-0.580) | 0.133 (0.048-0.234) | 0.356 (0.244-0.471) | 0.337 (0.226-0.457) | <0.001 |
| Full-scale data training (n=30) | 0.781 (0.714-0.836) | 0.760 (0.681-0.827) | 0.754 (0.696-0.808) | **0.785 (0.726-0.838)** | 0.756 (0.694-0.812) | 0.358 |
| **Pancreatic cyst** | | | | | | |
| No. of Few-shot Samples† | | | | | | |
| 1 | **0.465 (0.382-0.541)** | 0.077 (0.043-0.118) | 0.099 (0.054-0.153) | 0.056 (0.022-0.100) | 0.055 (0.023-0.092) | <0.001 |
| 2 | **0.460 (0.375-0.546)** | 0.104 (0.054-0.156) | 0.130 (0.074-0.197) | 0.081 (0.040-0.132) | 0.075 (0.040-0.118) | <0.001 |
| 4 | **0.515 (0.428-0.609)** | 0.203 (0.138-0.277) | 0.261 (0.178-0.352) | 0.179 (0.109-0.252) | 0.162 (0.099-0.229) | <0.001 |
| 8 | **0.641 (0.552-0.710)** | 0.220 (0.158-0.285) | 0.458 (0.389-0.534) | 0.187 (0.119-0.253) | 0.175 (0.115-0.241) | <0.001 |
| 16 | **0.707 (0.636-0.768)** | 0.308 (0.239-0.378) | 0.497 (0.412-0.579) | 0.092 (0.048-0.140) | 0.082 (0.044-0.125) | <0.001 |
| Full-scale data training (n=61) | **0.722 (0.653-0.788)** | 0.680 (0.599-0.751) | 0.637 (0.555-0.712) | 0.715 (0.641-0.779) | 0.682 (0.606-0.750) | 0.020 |
| **Kidney cyst** | | | | | | |
| No. of Few-shot Samples† | | | | | | |
| 1 | **0.161 (0.132-0.192)** | 0.045 (0.030-0.061) | 0.048 (0.033-0.066) | 0.056 (0.039-0.076) | 0.054 (0.038-0.071) | <0.001 |
| 2 | **0.137 (0.108-0.167)** | 0.013 (0.006-0.021) | 0.010 (0.004-0.019) | 0.000 (0.000-0.000) | 0.005 (0.004-0.007) | <0.001 |
| 4 | **0.141 (0.114-0.171)** | 0.004 (0.001-0.009) | 0.000 (0.000-0.000) | 0.000 (0.000-0.000) | 0.005 (0.004-0.006) | <0.001 |
| 8 | **0.102 (0.077-0.127)** | 0.000 (0.000-0.000) | 0.000 (0.000-0.000) | 0.000 (0.000-0.000) | 0.005 (0.004-0.007) | <0.001 |
| 16 | **0.289 (0.252-0.329)** | 0.000 (0.000-0.000) | 0.000 (0.000-0.000) | 0.000 (0.000-0.000) | 0.005 (0.004-0.006) | <0.001 |
| Full-scale data training (n=489) | **0.660 (0.622-0.696)** | 0.641 (0.607-0.677) | 0.616 (0.582-0.649) | 0.624 (0.589-0.655) | 0.598 (0.562-0.630) | <0.001 |

| **Kidney stone** | | | | | | |
|---|---|---|---|---|---|---|
| No. of Few-shot Samples† | | | | | | |
| 1 | **0.464**<br>**(0.350-0.573)** | 0.000<br>(0.000-0.000) | 0.000<br>(0.000-0.000) | 0.000<br>(0.000-0.000) | 0.002<br>(0.000-0.005) | <0.001 |
| 2 | **0.678**<br>**(0.622-0.730)** | 0.000<br>(0.000-0.000) | 0.000<br>(0.000-0.000) | 0.000<br>(0.000-0.000) | 0.005<br>(0.002-0.009) | <0.001 |
| 4 | **0.780**<br>**(0.733-0.819)** | 0.000<br>(0.000-0.000) | 0.000<br>(0.000-0.000) | 0.000<br>(0.000-0.000) | 0.006<br>(0.002-0.011) | <0.001 |
| 8 | **0.636**<br>**(0.502-0.752)** | 0.000<br>(0.000-0.000) | 0.000<br>(0.000-0.000) | 0.000<br>(0.000-0.000) | 0.005<br>(0.001-0.009) | <0.001 |
| 16 | **0.822**<br>**(0.799-0.844)** | 0.004<br>(0.000-0.011) | 0.000<br>(0.000-0.000) | 0.000<br>(0.000-0.000) | 0.005<br>(0.001-0.009) | <0.001 |
| Full-scale data training (n=30) | **0.777**<br>**(0.713-0.820)** | 0.755<br>(0.682-0.809) | 0.768<br>(0.707-0.814) | 0.755<br>(0.675-0.823) | 0.724<br>(0.638-0.789) | 0.261 |
| **Brain tumour (MRI)** | | | | | | |
| No. of Few-shot Samples† | | | | | | |
| 1 | **0.310**<br>**(0.288-0.333)** | 0.303<br>(0.281-0.325) | 0.312<br>(0.287-0.337) | 0.288<br>(0.265-0.309) | 0.253<br>(0.231-0.273) | 0.208 |
| 2 | **0.344**<br>**(0.318-0.372)** | 0.290<br>(0.269-0.311) | 0.310<br>(0.287-0.333) | 0.270<br>(0.247-0.292) | 0.240<br>(0.219-0.264) | <0.001 |
| 4 | **0.438**<br>**(0.415-0.462)** | 0.376<br>(0.353-0.399) | 0.403<br>(0.376-0.428) | 0.348<br>(0.324-0.370) | 0.315<br>(0.293-0.336) | <0.001 |
| 8 | **0.465**<br>**(0.441-0.488)** | 0.420<br>(0.397-0.443) | 0.450<br>(0.425-0.475) | 0.423<br>(0.399-0.447) | 0.386<br>(0.362-0.409) | 0.011 |
| 16 | **0.504**<br>**(0.479-0.528)** | 0.447<br>(0.425-0.471) | 0.468<br>(0.443-0.490) | 0.457<br>(0.434-0.483) | 0.418<br>(0.393-0.440) | <0.001 |
| **Liver tumour (MRI)** | | | | | | |
| No. of Few-shot Samples† | | | | | | |
| 1 | **0.267**<br>**(0.219-0.315)** | 0.211<br>(0.173-0.255) | 0.237<br>(0.185-0.287) | 0.213<br>(0.171-0.259) | 0.181<br>(0.141-0.227) | 0.030 |
| 2 | **0.339**<br>**(0.282-0.397)** | 0.225<br>(0.185-0.269) | 0.232<br>(0.185-0.291) | 0.193<br>(0.152-0.236) | 0.166<br>(0.127-0.209) | <0.001 |
| 4 | **0.492**<br>**(0.422-0.555)** | 0.284<br>(0.233-0.341) | 0.341<br>(0.279-0.405) | 0.252<br>(0.204-0.307) | 0.232<br>(0.183-0.280) | <0.001 |
| 8 | **0.562**<br>**(0.495-0.628)** | 0.289<br>(0.245-0.337) | 0.430<br>(0.368-0.501) | 0.296<br>(0.244-0.343) | 0.264<br>(0.212-0.319) | <0.001 |
| 16 | **0.603**<br>**(0.532-0.672)** | 0.292<br>(0.238-0.339) | 0.469<br>(0.400-0.540) | 0.283<br>(0.235-0.326) | 0.244<br>(0.197-0.292) | <0.001 |

Non-parametric bootstrapping with 1,000 bootstrap replicates is employed for statistical analysis. The 95% CI is included in parentheses.

* *p*-values were calculated using a one-sided Wilcoxon test comparing the **top-performing model** and the second-best model for each task

† 2000 iterations.

**eTable 10: Mask and Report Annotation Efficiency of Different Experience Levels Radiologists across Lesion Types with and without PASTA-AID Assistance**

| | Liver Cancer | | Pancreatic cancer | | Kidney Cancer | |
|---|---|---|---|---|---|---|
| | T_mask | T_report | T_mask | T_report | T_mask | T_report |
| **Junior Radiologist 1** | | | | | | |
| Solo (n=50) | 590.3 (534.0, 646.6) | 197.4 (179.4, 215.5) | 341.8 (303.2, 380.3) | 175.7 (159.8, 191.7) | 286.1 (265.9, 306.3) | 137.4 (118.9, 156.0) |
| PASTA-AID (n=50) | 294.8 (239.9, 349.6) | 146.5 (133.2, 159.9) | 80.8 (74.8, 86.8) | 122.4 (111.4, 133.3) | 62.0 (51.0, 73.0) | 121.1 (103.6, 138.6) |
| **Junior Radiologist 2** | | | | | | |
| Solo (n=50) | 597.0 (551.3, 642.6) | 189.0 (171.3, 206.6) | 321.2 (287.3, 355.1) | 173.5 (160.3, 186.8) | 298.9 (278.2, 319.7) | 143.8 (127.5, 160.2) |
| PASTA-AID (n=50) | 299.1 (252.5, 345.8) | 147.1 (133.8, 160.3) | 81.9 (75.9, 87.9) | 108.0 (97.1, 118.9) | 65.5 (55.6, 75.5) | 116.1 (104.8, 127.4) |
| ***p*-value** | <0.001 | <0.001 | <0.001 | <0.001 | <0.001 | 0.006 |
| **Overall time reduction (%)** | 50.0 | 24.0 | 75.5 | 34.0 | 78.2 | 15.7 |
| **Senior Radiologist 1** | | | | | | |
| Solo (n=50) | 495.4 (451.0, 539.7) | 157.6 (138.7, 176.6) | 263.6 (232.7, 294.5) | 144.9 (135.9, 153.9) | 235.1 (220.3, 249.9) | 136.2 (125.1, 147.3) |
| PASTA-AID (n=50) | 251.2 (196.9, 305.6) | 126.6 (112.9, 140.2) | 67.9 (62.4, 73.5) | 103.7 (96.7, 110.7) | 59.0 (46.0, 72.1) | 92.5 (83.4, 101.6) |
| **Senior Radiologist 2** | | | | | | |
| Solo (n=50) | 477.4 (437.8, 517.0) | 152.8 (135.9, 169.8) | 238.6 (216.6, 260.7) | 135.6 (127.6, 143.6) | 226.8 (212.4, 241.1) | 142.5 (131.9, 153.1) |
| PASTA-AID (n=50) | 280.4 (239.9, 321.0) | 116.1 (101.6, 130.6) | 70.6 (65.1, 76.0) | 111.5 (102.7, 120.4) | 58.7 (46.2, 71.2) | 84.5 (75.7, 93.3) |
| ***p-value**** | <0.001 | <0.001 | <0.001 | <0.001 | <0.001 | <0.001 |
| **Overall time reduction (%)** | 45.3 | 21.8 | 72.4 | 23.3 | 74.5 | 36.5 |

T_mask: mask annotation time (seconds), T_report: report generation time (seconds).

Time-related data within each cell are presented as mean (95% CI).

* *p*-values were calculated using a linear mixed-effects model with doctor as a random intercept.

**eTable 11: Performance Improvement of Junior Radiologists with PASTA-AID Assistance. Annotations from Senior Radiologists are set as the Ground Truth.**

| | Liver Cancer | | Pancreatic cancer | | Kidney Cancer | |
|---|---|---|---|---|---|---|
| | Dice | Accuracy | Dice | Accuracy | Dice | Accuracy |
| **Junior Radiologist 1** | | | | | | |
| Solo (n=50) | 0.740 (0.677, 0.803) | 0.940 (0.914, 0.966) | 0.740 (0.649, 0.832) | 0.908 (0.875, 0.941) | 0.752 (0.704, 0.799) | 0.932 (0.902, 0.962) |
| PASTA-AID (n=50) | 0.805 (0.744, 0.866) | 0.948 (0.920, 0.976) | 0.850 (0.784, 0.917) | 0.920 (0.890, 0.950) | 0.818 (0.763, 0.872) | 0.944 (0.918, 0.970) |
| **Junior Radiologist 2** | | | | | | |
| Solo (n=50) | 0.702 (0.628, 0.776) | 0.928 (0.898, 0.958) | 0.764 (0.703, 0.826) | 0.924 (0.894, 0.954) | 0.715 (0.659, 0.772) | 0.920 (0.888, 0.952) |
| PASTA-AID (n=50) | 0.813 (0.740, 0.887) | 0.924 (0.892, 0.956) | 0.832 (0.777, 0.886) | 0.936 (0.909, 0.963) | 0.826 (0.771, 0.880) | 0.916 (0.883, 0.949) |
| Performance Improvement (%) | 12.2 | 0.2 | 11.8 | 1.3 | 12.0 | 0.4 |
| ***p*-value** | **0.011** | 0.890 | **0.017** | 0.422 | **0.001** | 0.790 |

Dice: image overlap ratio, acc: accuracy of structured reports.

Dice scores for image overlap and accuracy rates of structured reports within each cell are presented as mean (95% CI).

* *p*-values were calculated using a linear mixed-effects model with doctor as a random intercept.

## eTable 12: Accuracy and F1-Scores of Various Models in Structured Report Generation

| | PASTA | UNet | Models Genesis | SuPerM | *p*-value* |
|---|---|---|---|---|---|
| **Shape** | | | | | |
| ACC | **0.849 (0.831-0.866)** | 0.789 (0.769-0.809) | 0.660 (0.636-0.681) | 0.788 (0.767-0.807) | <0.001 |
| F1-score | **0.781 (0.738-0.818)** | 0.688 (0.645-0.730) | 0.199 (0.194-0.203) | 0.742 (0.701-0.775) | 0.017 |
| **Invasion** | | | | | |
| ACC | **0.815 (0.797-0.833)** | 0.793 (0.772-0.815) | 0.754 (0.732-0.775) | 0.810 (0.790-0.830) | 0.319 |
| F1-score | 0.723 (0.699-0.749) | 0.700 (0.672-0.728) | 0.430 (0.422-0.437) | **0.732 (0.706-0.759)** | 0.730 |
| **Density** | | | | | |
| ACC | **0.726 (0.702-0.746)** | 0.689 (0.667-0.711) | 0.529 (0.503-0.553) | 0.637 (0.612-0.662) | <0.001 |
| F1-score | **0.684 (0.657-0.706)** | 0.644 (0.617-0.668) | 0.334 (0.315-0.352) | 0.589 (0.561-0.615) | <0.001 |
| **Heterogeneity** | | | | | |
| ACC | **0.903 (0.889-0.917)** | 0.871 (0.854-0.888) | 0.752 (0.729-0.771) | 0.829 (0.810-0.848) | <0.001 |
| F1-score | **0.868 (0.849-0.887)** | 0.825 (0.802-0.847) | 0.429 (0.422-0.435) | 0.762 (0.735-0.787) | <0.001 |
| **Surface** | | | | | |
| ACC | **0.891 (0.875-0.907)** | 0.863 (0.845-0.880) | 0.722 (0.702-0.744) | 0.811 (0.791-0.830) | <0.001 |
| F1-score | **0.863 (0.843-0.883)** | 0.825 (0.802-0.847) | 0.419 (0.413-0.427) | 0.754 (0.729-0.779) | <0.001 |

Non-parametric bootstrapping with 1,000 bootstrap replicates is employed for statistical analysis. The 95% CI is included in parentheses.

* *p*-values were calculated using a one-sided permutation test with 10,000 permutations comparing the **top-performing model** and the second-best model for each task

## eTable 13: Comparison of Tumour Staging and Survival Prediction AUC Values.

| | PASTA | Models Genesis | SuPerM | FMCIB | *p*-value* |
|---|---|---|---|---|---|
| **Tumour staging** | | | | | |
| Gastric cancer | **0.770 (0.636-0.858)** | 0.548 (0.456-0.656) | 0.618 (0.520-0.751) | 0.741 (0.593-0.851) | 0.031 |
| Rectal cancer | **0.738 (0.643-0.846)** | 0.646 (0.617-0.683) | 0.613 (0.421-0.804) | 0.645 (0.461-0.779) | 0.031 |
| Bladder cancer | **0.855 (0.750-0.950)** | 0.633 (0.489-0.800) | 0.648 (0.500-0.889) | 0.689 (0.646-0.756) | 0.031 |
| **Survival prediction** | | | | | |
| Lung cancer | **0.700 (0.595-0.779)** | 0.650 (0.562-0.701) | 0.687 (0.597-0.730) | 0.617 (0.591-0.660) | 0.156 |
| Gastric cancer | **0.660 (0.525-0.784)** | 0.546 (0.480-0.652) | 0.508 (0.344-0.647) | 0.626 (0.481-0.763) | 0.031 |
| Rectal cancer | **0.759 (0.718-0.858)** | 0.626 (0.468-0.750) | 0.714 (0.656-0.790) | 0.713 (0.613-0.805) | 0.094 |
| Bladder cancer | **0.878 (0.750-1.000)** | 0.844 (0.722-1.000) | 0.838 (0.700-1.000) | 0.843 (0.750-1.000) | 0.500 |

The values in parentheses represent the minimum and maximum AUC values obtained across the 5-fold cross-validation.

* *p*-values were calculated using a one-sided Wilcoxon test comparing the **top-performing model** and the second-best model for each task

**eTable 14: CT modalities selected as templates for simulating each lesion in PASTA-Gen**

| Lesion | Modality of template CT |
|---|---|
| Lung tumour | Non-contrast CT |
| Liver tumour | Enhanced CT |
| Gallbladder cancer | Enhanced CT |
| Pancreatic tumour | Enhanced CT |
| Esophageal Cancer | Enhanced CT |
| Gastric cancer | Enhanced CT |
| Colorectal cancer | Enhanced CT |
| Kidney tumour | Enhanced CT |
| Bladder cancer | Enhanced CT |
| Bone metastasis | Enhanced CT |
| Liver cyst | Enhanced CT & Non-contrast CT |
| Gallstone | Enhanced CT & Non-contrast CT |
| Pancreatic cyst | Enhanced CT |
| Kidney cyst | Enhanced CT |
| Kidney stone | Non-contrast CT |

**eTable 15: Class name and value in PASTA-Gen-30K**

| Lesion | Value |
|---|---|
| Lung | 1 |
| Liver | 2 |
| Gallbladder | 3 |
| Pancreas | 4 |
| Esophagus | 5 |
| Stomach | 6 |
| Colon & rectal | 7 |
| Kidney | 8 |
| Bladder | 9 |
| Bone | 10 |
| Lung tumour | 11 |
| Liver tumour | 12 |
| Gallbladder cancer | 13 |
| Pancreatic tumour | 14 |
| Esophageal Cancer | 15 |
| Gastric cancer | 16 |
| Colorectal cancer | 17 |
| Kidney tumour | 18 |
| Bladder cancer | 19 |
| Bone metastasis | 20 |
| Liver cyst | 21 |
| Gallstone | 22 |
| Pancreatic cyst | 23 |
| Kidney cyst | 24 |
| Kidney stone | 25 |

## eReferences

1. Nobel JM, van Geel K, Robben SG. Structured reporting in radiology: a systematic review to explore its potential. *European radiology* 2022: 1-18.
2. Wasserthal J, Breit H-C, Meyer MT, et al. TotalSegmentator: robust segmentation of 104 anatomic structures in CT images. *Radiology: Artificial Intelligence* 2023; **5**(5).
3. Ho J, Jain A, Abbeel P. Denoising diffusion probabilistic models. *Advances in neural information processing systems* 2020; **33**: 6840-51.
4. Yu Y, Chen H, Zhang Z, et al. CT Synthesis with Conditional Diffusion Models for Abdominal Lymph Node Segmentation. *arXiv preprint arXiv:240317770* 2024.
5. Loshchilov I, Hutter F. Decoupled weight decay regularization. *arXiv preprint arXiv:171105101* 2017.
6. Çiçek Ö, Abdulkadir A, Lienkamp SS, Brox T, Ronneberger O. 3D U-Net: learning dense volumetric segmentation from sparse annotation. Medical Image Computing and Computer-Assisted Intervention–MICCAI 2016: 19th International Conference, Athens, Greece, October 17-21, 2016, Proceedings, Part II 19; 2016: Springer; 2016. p. 424-32.
7. Ulyanov D. Instance normalization: The missing ingredient for fast stylization. *arXiv preprint arXiv:160708022* 2016.
8. Isensee F, Jaeger PF, Kohl SA, Petersen J, Maier-Hein KH. nnU-Net: a self-configuring method for deep learning-based biomedical image segmentation. *Nature methods* 2021; **18**(2): 203-11.
9. Tu T, Azizi S, Driess D, et al. Towards generalist biomedical AI. *NEJM AI* 2024; **1**(3): AIoa2300138.
10. Ma J, Zhang Y, Gu S, et al. Unleashing the strengths of unlabelled data in deep learning-assisted pan-cancer abdominal organ quantification: the FLARE22 challenge. *The Lancet Digital Health* 2024; **6**(11): e815-e26.
11. Milletari F, Navab N, Ahmadi S-A. V-net: Fully convolutional neural networks for volumetric medical image segmentation. 2016 fourth international conference on 3D vision (3DV); 2016: Ieee; 2016. p. 565-71.
12. Zhou Z, Sodha V, Pang J, Gotway MB, Liang J. Models genesis. *Medical image analysis* 2021; **67**: 101840.
13. Setio AAA, Traverso A, De Bel T, et al. Validation, comparison, and combination of algorithms for automatic detection of pulmonary nodules in computed tomography images: the LUNA16 challenge. *Medical image analysis* 2017; **42**: 1-13.
14. Li W, Yuille A, Zhou Z. How well do supervised 3d models transfer to medical imaging tasks? *arXiv preprint arXiv:250111253* 2025.
15. Li W, Qu C, Chen X, et al. Abdomenatlas: A large-scale, detailed-annotated, & multi-center dataset for efficient transfer learning and open algorithmic benchmarking. *Medical Image Analysis* 2024; **97**: 103285.
16. Pai S, Bontempi D, Hadzic I, et al. Foundation model for cancer imaging biomarkers. *Nature machine intelligence* 2024; **6**(3): 354-67.
17. He K, Zhang X, Ren S, Sun J. Deep residual learning for image recognition. Proceedings of the IEEE conference on computer vision and pattern recognition; 2016; 2016. p. 770-8.
18. Yan K, Wang X, Lu L, Summers RM. DeepLesion: automated mining of large-scale lesion annotations and universal lesion detection with deep learning. *Journal of medical imaging* 2018; **5**(3): 036501-.
19. Antonelli M, Reinke A, Bakas S, et al. The medical segmentation decathlon. *Nature communications* 2022; **13**(1): 4128.
20. Heller N, Isensee F, Maier-Hein KH, et al. The state of the art in kidney and kidney tumor segmentation in contrast-enhanced CT imaging: Results of the KiTS19 challenge. *Medical image analysis* 2021; **67**: 101821.
21. Aerts HJ, Velazquez ER, Leijenaar RT, et al. Decoding tumour phenotype by noninvasive imaging using a quantitative radiomics approach. *Nature communications* 2014; **5**(1): 4006.
22. Kirk S, Lee Y, Lucchesi F, et al. The Cancer Genome Atlas Urothelial Bladder Carcinoma Collection (TCGA-BLCA)(Version 8)[Data set]. *The Cancer Imaging Archive* 2016; **10**: K9.
23. Quinton F, Popoff R, Presles B, et al. A tumour and liver automatic segmentation (atlas) dataset on contrast-enhanced

magnetic resonance imaging for hepatocellular carcinoma. *Data* 2023; **8**(5): 79.